\documentclass[12pt]{article}
\pdfoutput=1
\usepackage{putex}
\usepackage{graphicx}
\usepackage{caption}
\usepackage{subcaption}
\usepackage{epstopdf}
\usepackage{enumerate}
\usepackage{cite}
\usepackage{youngtab}
\usepackage{braket}
\usepackage{bbm}
\usepackage{tensor}
\usepackage{slashed}
\usepackage[aligntableaux=center]{ytableau}
\usepackage{multirow}
\newcommand{\abs}[1]{\left\lvert #1 \right\rvert}

\newcommand{\vphi}{\varphi}

\newcommand {\be} {\begin {equation}}
\newcommand {\ee} {\end {equation}}

\newcommand {\bes} {\begin {equation*}}
\newcommand {\ees} {\end {equation*}}

\newcommand{\es}[2] {\begin{equation} \label{#1} \begin{split} #2 \end{split} \end{equation}}

\newcommand{\Z}{\mathbb{Z}}

\newcommand{\R}{\mathbb{R}}
\newcommand{\C}{\mathbb{C}}

\newcommand{\cO}{{\cal O}}

\newcommand{\beq}{\begin{equation}}
\newcommand{\eeq}{\end{equation}}

\begin{document}

\preprint{PUPT-2541}

\institution{HU}{Jefferson Physical Laboratory, Harvard University, Cambridge, MA 02138, USA }
\institution{PU}{Joseph Henry Laboratories, Princeton University, Princeton, NJ 08544, USA}

\title{Solving M-theory with the Conformal Bootstrap}

\authors{Nathan B.~Agmon,\worksat{\HU} Shai M.~Chester,\worksat{\PU} and Silviu S.~Pufu\worksat{\PU}}

\abstract{
We use the conformal bootstrap to perform a precision study of 3d maximally supersymmetric ($\mathcal{N}=8$) SCFTs that describe the IR physics on $N$ coincident M2-branes placed either in flat space or at a $\C^4/\Z_2$ singularity.  
First, using the explicit Lagrangians of ABJ(M) \cite{Aharony:2008ug,Aharony:2008gk} and recent supersymmetric localization results, we calculate certain half and quarter-BPS OPE coefficients, both exactly at small $N$, and approximately in a large $N$ expansion that we perform to all orders in $1/N$.  Comparing these values with the numerical bootstrap bounds leads us to conjecture that some of these theories obey an OPE coefficient minimization principle.  
We then use this conjecture as well as the extremal functional method to reconstruct the first few low-lying scaling dimensions and OPE coefficients for both protected and unprotected multiplets that appear in the OPE of two stress tensor multiplets for all values of $N$.   We also calculate the half and quarter-BPS operator OPE coefficients in the $SU(2)_k \times SU(2)_{-k}$ BLG theory for all values of the Chern-Simons coupling $k$, and show that generically they do not obey the same OPE coefficient minimization principle.
}
\date{\today}

\maketitle

\tableofcontents

\section{Introduction and summary}

The conformal bootstrap \cite{Polyakov:1974gs,Ferrara:1973yt, Mack:1975jr } can be used to place rigorous bounds on scaling dimensions and operator product expansion (OPE) coefficients of operators that appear in the conformal block decomposition of a given four-point function or of a given system of four-point functions  \cite{Rattazzi:2008pe}.   Generally, these bounds get more and more stringent as one explores a larger and larger set of crossing symmetry constraints.\footnote{For a numerical implementation, one has to truncate the set of crossing constraints of given four-point function(s) to a finite number of  equations controlled by a parameter $\Lambda$.  It is customary to perform this truncation by only considering derivatives of the crossing equations at the crossing-symmetric point whose total order is at most $\Lambda$.}  Quite remarkably, in certain cases, the bounds include small islands of allowed regions in theory space that contain well-known CFTs \cite{Kos:2014bka,Kos:2015mba,Kos:2016ysd}.\footnote{For other examples where the numerical bootstrap was used to obtain bounds on the OPE coefficients and/or scaling dimensions, see \cite{Rychkov:2009ij,Poland:2010wg,Vichi:2011ux,Poland:2011ey,Liendo:2012hy,Kos:2013tga,El-Showk:2013nia,Gaiotto:2013nva,Chester:2014gqa,Beem:2014zpa,Chester:2015qca,Iliesiu:2015qra,Poland:2015mta,Lemos:2015awa, Chester:2015lej,Lin:2015wcg,Chester:2016wrc,Lemos:2016xke,Iliesiu:2017nrv,Cornagliotto:2017dup,Li:2017ddj,Dymarsky:2017xzb,Chester:2017vdh,Chang:2017xmr,Chang:2017cdx,Lin:2016gcl}.}  For a generic point within the allowed region\footnote{Here, we mean the limit of the allowed region as we remove the cutoff that controls the truncation of the crossing equation to a finite number.} formed by these bounds, there are generally many different solutions to the crossing equations obeying unitarity constraints.   At the boundary of the allowed region, however, there is believed to be a unique such solution, which can then be used to read off the CFT data (scaling dimensions and OPE coefficients) that enters the conformal block decomposition of the given four-point function(s).  This solution can be found, for instance, using the extremal functional method of \cite{Poland:2010wg, ElShowk:2012hu, El-Showk:2014dwa}.  If we have reasons to believe that a known CFT lives on this boundary, we can therefore potentially determine at least part of its CFT data.  

A notable application of this method is to the 3d Ising model. In \cite{El-Showk:2014dwa}, it was argued that the critical Ising model has the minimal value of the stress tensor coefficient $c_T$ (to be defined more precisely shortly) in the space of possible 3d CFTs with $\mathbb{Z}_2$ symmetry, and thus it is believed to sit at the boundary of the region of allowed values of $c_T$.  Reconstructing the corresponding unique solution of the crossing equations using the extremal functional method, one can then read off all low-lying CFT data in the critical Ising model.  See \cite{Simmons-Duffin:2016wlq, El-Showk:2016mxr,Bobev:2015jxa,Bobev:2015vsa} for other cases where this method was applied.

In this paper, we will apply the extremal functional method to maximally supersymmetric ($\mathcal{N}=8$) superconformal field theories (SCFTs) in 3d. To do so, we would first need to argue that the 3d ${\cal N} = 8$ SCFTs of interest to us saturate some numerical bounds on OPE coefficients or scaling dimensions.  Indeed, using supersymmetric localization we will be able to calculate the values of certain OPE coefficients that we can also bound using the numerical bootstrap technique.  As we will show, for ${\cal N} = 8$ SCFTs with holographic duals, these OPE coefficients come very close to saturating the bootstrap bounds obtained numerically.  We conjecture that these OPE coefficients for some of these theories precisely saturate the bootstrap bounds in the limit of very precise numerics.

Let us now provide more background and summarize our results.  There are only a few known infinite families of ${\cal N}=8$ SCFTs, and they can all be realized (in ${\cal N } = 3$ SUSY notation) as Chern-Simons (CS) theories with a product gauge group $G_1\times G_2$ coupled to two matter hypermultiplets transforming in the bifundamental representation. These families are:  the $SU(2)_k \times SU(2)_{-k}$ reformulation \cite{VanRaamsdonk:2008ft, Bandres:2008vf} of the theories of Bagger-Lambert-Gustavsson (BLG) \cite{Bagger:2007vi,Bagger:2007jr,Bagger:2006sk, Gustavsson:2007vu}, which are indexed by an arbitrary integer Chern-Simons level $k$;  the $U(N)_k \times U(N)_{-k}$ theories of Aharony-Bergman-Jafferis-Maldacena (ABJM) \cite{Aharony:2008ug}, which are labeled by the integer $N$ and $k = 1, 2$; and the $U(N+1)_2 \times U(N)_{-2}$ theories \cite{Bashkirov:2011pt} of Aharony-Bergman-Jafferis (ABJ) \cite{Aharony:2008gk}, which are labeled by the integer $N$. We will denote these theories as BLG$_k$,\footnote{There are actually two BLG type theories, with gauge groups $SU(2)_k \times SU(2)_{-k}$ and $(SU(2)_k \times SU(2)_{-k})/\mathbb{Z}_2$. The difference between them will not matter for the four-point function that we consider in this paper, so we denote both by BLG$_k$.} ABJM$_{N,k}$, and ABJ$_N$, respectively. When $N=1$, the ABJM$_{1, 1}$ theory describes a free SCFT equivalent to the theory of eight massless real scalars and eight Majorana fermions.  When $N>1$, ABJM$_{N, 1}$ flows to two decoupled SCFTs in the infrared: a free SCFT isomorphic to ABJM$_{1, 1}$, and a strongly coupled interacting SCFT\@. The ABJ(M) theories can be interpreted as effective theories on $N$ coincident M2-branes placed at a $\mathbb{C}^4/\mathbb{Z}_k$ singularity in the transverse direction, so that when $N\to\infty$ they contain a sector described by weakly coupled supergravity. The BLG theories, in contrast, do not have a known M-theory interpretation except when $k\leq4$, in which case they are dual to an ABJ(M) theory \cite{Agmon:2017lga,Lambert:2010ji,Bashkirov:2011pt,Bashkirov:2012rf}.

$\mathcal{N}=8$ SCFTs were first studied using the conformal bootstrap in \cite{Chester:2014fya, Chester:2014mea}, in which upper and lower bounds were placed on the CFT data that enters the conformal block expansion of the four-point function of the stress-tensor multiplet. These bounds were computed as a function of $c_T$, which is defined as the coefficient appearing in the two-point function of the canonically-normalized stress-tensor, 
 \es{CanStress}{
  \langle T_{\mu\nu}(\vec{x}) T_{\rho \sigma}(0) \rangle = \frac{c_T}{64} \left(P_{\mu\rho} P_{\nu \sigma} + P_{\nu \rho} P_{\mu \sigma} - P_{\mu\nu} P_{\rho\sigma} \right) \frac{1}{16 \pi^2 \vec{x}^2} \,, \qquad P_{\mu\nu} \equiv \eta_{\mu\nu} \nabla^2 - \partial_\mu \partial_\nu \,.
 }
This coefficient can be computed exactly using supersymmetric localization for any ${\cal N} \geq 2$ SCFT with a Lagrangian description (see \cite{Closset:2012ru} and \cite{Imamura:2011wg}).  In \eqref{CanStress}, $c_T$ is normalized such that it equals $1$ for a (non-supersymmetric) free massless real scalar or a free massless Majorana fermion.   Thus, $c_T = 16$ for the free ${\cal N} = 8$ theory of eight massless real scalars and eight massless Majorana fermions (equivalent to ABJM$_{1, 1}$), and 
\es{cTlargeN}{
  c_T \approx \frac{64}{3\pi}\sqrt{2k}N^{3/2}
}
for ABJ(M) theory at large $N$.  By varying $c_T$ over the range $c_T \in [16, \infty)$, one could obtain non-perturbative information about M-theory, albeit only in the form of bounds.

In \cite{Agmon:2017lga}, the OPE coefficients of the half and quarter-BPS operators that appear in the stress-tensor four-point function were computed for BLG$_3$ and the interacting sector of ABJM$_{3,1}.$\footnote{These theories are conjectured to be dual \cite{Agmon:2017lga}, so the OPE coefficients are in fact identical for both theories.}  This calculation was made possible due to the observation that ${\cal N} = 4$ SCFTs (of which ${\cal N} = 8$ SCFTs are a particular class) contain 1d topological sectors \cite{Chester:2014mea,Beem:2016cbd,Beem:2013sza} that can be accessed explicitly using supersymmetric localization \cite{Dedushenko:2016jxl}.  The OPE coefficients were found to saturate the lower bounds obtained numerically in \cite{Chester:2014mea}, just as $c_T$ saturated the lower bounds in the Ising model, which suggests that the extremal functional approach can be used in this case.

In this work, we generalize the computation of \cite{Agmon:2017lga} to all $\mathcal{N}=8$ SCFTs mentioned above.  Instead of working directly in the 1d topological theory obtained from supersymmetric localization \cite{Dedushenko:2016jxl}, we argue that one can relate certain integrated correlators in the 1d theory to derivatives of the partition function of an ${\cal N} = 4$-preserving mass deformation of the SCFT on $S^3$.   For each theory, this mass-deformed $S^3$ partition function can be expressed as a matrix integral using the results of Kapustin, Willet, and Yaakov \cite{Kapustin:2009kz}.  For BLG$_k$, the matrix integral can be computed exactly for all $k$.  For ABJ(M) theory, the corresponding integrals can be computed either exactly at small $N$, or to all orders in the $1/N$ expansion using the Fermi gas methods in \cite{Nosaka:2015iiw}.  From the mass-deformed partition function, one can then determine the integrated four-point function in the 1d theory, and from it, as well as from crossing symmetry, one can extract the OPE coefficients of interest.

The analytic expressions for the OPE coefficients can then be compared to the numerical bootstrap bounds. We find that the lower bound is close to being saturated by these OPE coefficients in ABJ$_N$ and the interacting sector of ABJM$_{N,1}$ for all $N$, in ABJM$_{N,2}$ for all $N\neq2$, and in BLG$_k$ only when $k=3,4$. These are exactly the cases when these theories have a unique stress tensor and have M-theory duals.\footnote{$(SU(2)\times SU(2))/\mathbb{Z}_2$ BLG is dual to ABJM$_{2,1}$, which is a product of a free and interacting theory. ABJM$_{2,2}$ is dual to $SU(2)\times SU(2)$ BLG and $(SU(2)\times SU(2))/\mathbb{Z}_2$ BLG is dual to two copies of ABJ$_{1}$. Both BLG theories have the same stress tensor four-point function, which makes ABJM$_{2,2}$ a product theory in this sector. For $k=3,4$, BLG$_k$ is dual to the interacting sector ABJM$_{3,1}$ and ABJ$_2$, respectively.} The formulae for these OPE coefficients become $k$-dependent beyond order $1/c_T$, so only one of these curves could saturate the numerical bounds at infinite precision, but the analytic expressions are too similar to distinguish between them at the current level of numerical precision. This motivates our conjecture that all $\mathcal{N}=8$ theories with holographic duals saturate the bootstrap bounds at large $c_T$, and that at least one of them saturates the bounds for all $c_T$, so that we can apply the extremal functional method to it.  Using this technique, we read off all the low-lying CFT data in the stress-tensor four-point function, including that of unprotected operators, for theories that are dual to M-theory. This provides a complete numerical non-perturbative description of M-theory for the operators in this sector.

The rest of this paper is organized as follows. In Section~\ref{4point} we review the conformal block decomposition of the four-point function of the scalar operator at the bottom of the ${\cal N} = 8$ stress tensor multiplet.  In Section~\ref{1d}, we explain our method for computing the OPE coefficients, and we perform this computation for BLG and ABJ(M) theory.  In Section~\ref{min}, we present our evidence for the conjecture that holographic theories saturate the bootstrap bounds on the OPE coefficients we computed in Section~\ref{1d}.  In Section~\ref{spec}, we use the extremal functional method to read off all the low-lying CFT data for theories that saturate the bootstrap bounds.  Finally, in Section~\ref{disc}, we end with a discussion of our results and of future directions.

\section{Four-point function of stress-tensor}
\label{4point}

Let us begin by reviewing some general properties of the four-point function of the stress-tensor multiplet in an $\mathcal{N}=8$ SCFT, and of the constraints imposed by the $\mathfrak{osp}(8|4)$ superconformal algebra (for more details, the reader is referred to e.g. \cite{Minwalla:1997ka,Bhattacharya:2008zy,Dolan:2008vc}).

Unitary irreps of $\mathfrak{osp}(8|4)$ are specified by the quantum numbers of their bottom component, namely by its scaling dimension $\Delta$, Lorentz spin $j$, and $\mathfrak{so}(8)$ R-symmetry irrep with Dynkin labels $[a_1\, a_2 \, a_3 \, a_4]$, as well as by various shortening conditions.  There are twelve different types of multiplets that we list in Table~\ref{Multiplets}.\footnote{The convention we use in defining these multiplets is that the supercharges transform in the ${\bf 8}_v = [1000]$ irrep of $\mathfrak{so}(8)_R$.}  
\begin{table}[http]
\begin{center}
\begin{tabular}{|l|c|c|c|c|}
\hline
 Type     & BPS    & $\Delta$             & Spin & $\mathfrak{so}(8)_R$  \\
 \hline 
 $(A,0)$ (long)      & $0$    & $\ge \Delta_0 + j+1$ & $j$  & $[a_1 a_2 a_3 a_4]$  \\
 $(A, 1)$  & $1/16$ & $\Delta_0 + j +1$    & $j$  & $[a_1 a_2 a_3 a_4]$  \\
 $(A, 2)$  & $1/8$  & $\Delta_0 + j +1$    & $j$  & $[0 a_2 a_3 a_4]$   \\
 $(A, 3)$  & $3/16$  & $\Delta_0 + j +1$    & $j$  & $[0 0 a_3 a_4]$     \\
 $(A, +)$  & $1/4$  & $\Delta_0 + j +1$    & $j$  & $[0 0 a_3 0]$       \\
 $(A, -)$  & $1/4$  & $\Delta_0 + j +1$    & $j$  & $[0 0 0 a_4]$       \\
 $(B, 1)$  & $1/8$  & $\Delta_0$           & $0$  & $[a_1 a_2 a_3 a_4]$ \\
 $(B, 2)$  & $1/4$  & $\Delta_0$           & $0$  & $[0 a_2 a_3 a_4]$   \\
 $(B, 3)$  & $3/8$  & $\Delta_0$                  & $0$  & $[0 0 a_3 a_4]$     \\
 $(B, +)$  & $1/2$  & $\Delta_0$           & $0$  & $[0 0 a_3 0]$       \\
 $(B, -)$  & $1/2$  & $\Delta_0$           & $0$  & $[0 0 0 a_4]$       \\
 conserved & $5/16$  & $j+1$                & $j$  & $[0 0 0 0]$         \\
 \hline
\end{tabular}
\end{center}
\caption{Multiplets of $\mathfrak{osp}(8|4)$ and the quantum numbers of their corresponding superconformal primary operator. The conformal dimension $\Delta$ is written in terms of $\Delta_0 \equiv a_1 + a_2 + (a_3 + a_4)/2$.  The Lorentz spin can take the values $j=0, 1/2, 1, 3/2, \ldots$.  Representations of the $\mathfrak{so}(8)$ R-symmetry are given in terms of the four $\mathfrak{so}(8)$ Dynkin labels, which are non-negative integers.}
\label{Multiplets}
\end{table}
There are two types of shortening conditions denoted by the $A$ and $B$ families.  The multiplet denoted by $(A, 0)$ is a long multiplet and does not obey any shortening conditions.  The other multiplets of type $A$ have the property that certain $\mathfrak{so}(2, 1)$ irreps of spin $j-1/2$ are absent from the product between the supercharges and the superconformal primary.  The multiplets of type $B$ have the property that certain $\mathfrak{so}(2, 1)$ irreps of spin $j \pm 1/2$ are absent from this product, and consequently, the multiplets of type $B$ are smaller.\footnote{This description is correct only when $j>0$.  When $j=0$, the definition of the multiplets also requires various conditions when acting on the primary with two supercharges.} 

The stress-tensor multiplet is of $(B, +)$ type,\footnote{Whether it is $(B, +)$ or $(B, -)$ is a matter of convention.  The two choices are related by the triality of $\mathfrak{so}(8)_R$.} with its superconformal primary being a dimension $1$ scalar operator transforming in the ${\bf 35}_c = [0020]$ irrep of $\mathfrak{so}(8)$.  Let us denote this superconformal primary by $\cO_{\text{Stress}, IJ}(\vec{x})$.  (The indices here are ${\bf 8}_c$ indices, and  $\cO_{\text{Stress}, IJ}(\vec{x})$ is a rank-two traceless symmetric tensor.)  In order to not carry around the $\mathfrak{so}(8)$ indices, it is convenient to contract them with an auxiliary polarization vector $Y^I$ that is constrained to be null $Y \cdot Y \equiv \sum_{I=1}^8 (Y^I)^2 = 0$, thus defining 
 \es{ODef}{
  \cO_\text{Stress}(\vec{x},Y) \equiv \cO_{\text{Stress}, IJ}(\vec{x}) Y^I Y^J \,.
 }

In the rest of this paper we will only consider the four-point function of $\cO_\text{Stress}(\vec{x},Y)$. Superconformal invariance implies that it takes the form
 \es{FourPointO}{
  \langle \cO_\text{Stress}(\vec{x}_1,Y_1) \cO_\text{Stress}(\vec{x}_2,Y_2)
   &\cO_\text{Stress}(\vec{x}_3,Y_3) \cO_\text{Stress}(\vec{x}_4,Y_4) \rangle \\
    &= \frac{( Y_1\cdot Y_2 )^2  ( Y_3\cdot Y_4 )^2 }
     {\abs{\vec{x}_{12}}^2\abs{\vec{x}_{34}}^2 }
     \sum_{{\cal M}\, \in\, \mathfrak{osp}(8|4) \text{ multiplets}} \lambda_\mathcal{M}^2 \mathcal{G_M}(u, v; \sigma, \tau) \,,
 }
where 
 \es{uvsigmatauDefs}{
  u \equiv \frac{\vec{x}_{12}^2 \vec{x}_{34}^2}{\vec{x}_{13}^2 \vec{x}_{24}^2} \,, \qquad
   v \equiv \frac{\vec{x}_{14}^2 \vec{x}_{23}^2}{\vec{x}_{13}^2 \vec{x}_{24}^2}  \,, \qquad
   \sigma\equiv\frac{(Y_1\cdot Y_3)(Y_2\cdot Y_4)}{(Y_1\cdot Y_2)(Y_3\cdot Y_4)}\,,\qquad \tau\equiv\frac{(Y_1\cdot Y_4)(Y_2\cdot Y_3)}{(Y_1\cdot Y_2)(Y_3\cdot Y_4)} \,,
 }
and the $\mathcal{G_M}$ correspond to the irreducible representations ${\cal M}$ of the superconformal algebra that appear in the OPE $\cO_\text{Stress} \times  \cO_\text{Stress}$. In Table~\ref{opemult}, we list the supermultiplets that may appear in this four-point function, following the constraints discussed in \cite{Ferrara:2001uj}. Since these are the only multiplets we will consider in this paper, we denote the short multiplets other than the stress-tensor as $(B,+)$ and $(B,2)$, the semi short multiplets as $(A,2)_j$ and $(A,+)_j$ where $j$ is the spin, and the long multiplet as $(A,0)_{j,n}$, where $n=0,1,\dots$ denotes the $n$th lowest multiplet with that spin---See the last column of Table~\ref{opemult}.  For explicit expressions for the functions ${\cal G_M}$, see \cite{Chester:2014fya}.

\begin{table}
\centering
\begin{tabular}{|c|c|r|c|c|}
\hline
Type    & $(\Delta,j)$     & $\mathfrak{so}(8)_R$ irrep  &spin $j$ & Name \\
\hline
$(B,+)$ &  $(2,0)$         & ${\bf 294}_c = [0040]$& $0$ & $(B, +)$ \\ 
$(B,2)$ &  $(2,0)$         & ${\bf 300} = [0200]$& $0$ & $(B, 2)$ \\
$(B,+)$ &  $(1,0)$         & ${\bf 35}_c = [0020]$ & $0$ & Stress \\
$(A,+)$ &  $(j+2,j)$       & ${\bf 35}_c = [0020]$ &even & $(A,+)_j$ \\
$(A,2)$ &  $(j+2,j)$       & ${\bf 28} = [0100]$ & odd & $(A,2)_j$ \\
$(A,0)$ &  $\Delta\ge j+1$ & ${\bf 1} = [0000]$ & even & $(A,0)_{j,n}$\\
\hline
\end{tabular}
\caption{The possible superconformal multiplets in the $\cO_\text{Stress} \times  \cO_\text{Stress}$ OPE\@.  The $\mathfrak{so}(3, 2) \oplus \mathfrak{so}(8)_R$ quantum numbers are those of the superconformal primary in each multiplet.}
\label{opemult}
\end{table}

Of particular importance will be the OPE coefficient which the stress tensor multiplet enters in the four-point function \eqref{FourPointO}.  In the conventions of \cite{Chester:2014fya}, if we normalize ${\cal O}_\text{Stress}$ such that the OPE coefficient of the identity operator is $\lambda_{\text{Id}} = 1$, then
 \es{cTRel}{
  \lambda_{\text{Stress}}^2 = \frac{256}{c_T} \,,
 }
where $c_T$ is the coefficient appearing in the two-point function \eqref{CanStress} of the canonically normalized stress tensor.  The $c_T$ coefficient was computed using supersymmetric localization in \cite{Chester:2014fya} for the known ${\cal N} = 8$ SCFTs with Lagrangian descriptions, and in the next section we will reproduce this result from a different calculation.

It is worth pointing out two limits in which the four-point function \eqref{FourPointO} is known exactly and one can extract all OPE coefficients.  The first limit is the free theory of eight real scalars $X_I$ and eight Majorana fermions mentioned in the Introduction.  The scalar ${\cal O}_{\text{Stress}, IJ}$ in this case is given by 
 \es{OIJFree}{
  {\cal O}_{\text{Stress}, IJ} = X_I X_J - \frac{\delta_{IJ}}{8} X_K X^K \,.
 }
%We can compute the four point function \eqref{FourPointO} explicitly in the free theory, where we can write the stress tensor operator superconformal primary in terms of the dimension one free multiplet superconformal primary as $ \cO^\text{free}_{\text{Stress}, IJ}(\vec{x}) = \cO_{\text{free}, I}(\vec{x})  \cO_{\text{free}, J}(\vec{x}) $. 
Performing Wick contractions with the propagator $\langle X_I(\vec{x}) X_J(0) \rangle = \frac{\delta_{IJ}}{4 \pi \abs{\vec{x}}}$, we then find that \eqref{FourPointO} equals:
\es{free4}{
 \frac{2}{(4 \pi)^4} \frac{( Y_1\cdot Y_2 )^2  ( Y_3\cdot Y_4 )^2 }
     {\abs{\vec{x}_{12}}^2\abs{\vec{x}_{34}}^2 }
    \left[1+u\sigma^2+\frac{u}{v}\tau^2+4\sqrt{u}\sigma+
    4\sqrt{\frac{u}{v}}\tau+4\frac{u}{\sqrt{v}}\sigma\tau\right]\,.
}
By comparing this to the conformal block expansion, we can read off the OPE coefficients listed in Table~\ref{Avalues}, where the scaling dimensions of the long multiplet are given by
\es{superFree}{
\Delta_{(A,0)_{j,n}}^{\text{free}}=j+\delta_{n,0}+2n\,,
}
with $n=0, 1, 2, \ldots$.  

Another limit in which we can compute \eqref{FourPointO}  explicitly is in the generalized free field theory (GFFT) where the dimension one operator $ \cO^\text{GFFT}_{\text{Stress}, IJ}(\vec{x})$ is treated as a generalized free field with two-point function $\langle {\cal O}_\text{Stress}(\vec{x}, Y_1)  {\cal O}_\text{Stress}(0, Y_2) \rangle = \frac{(Y_1 \cdot Y_2)^2}{\abs{x}^2}$.  The GFFT describes the $c_T\to\infty$, i.e. $\lambda^2_\text{Stress}=0$, limit of $\mathcal{N}=8$ theories.  Performing the Wick contractions, we then find
\es{mean4}{
\frac{( Y_1\cdot Y_2 )^2  ( Y_3\cdot Y_4 )^2 }
     {\abs{\vec{x}_{12}}^2\abs{\vec{x}_{34}}^2 }
    \left[1+u\sigma^2+\frac{u}{v}\tau^2\right]\,.
}
By comparing this to \eqref{FourPointO}, we can read off the OPE coefficients listed in Table \ref{Avalues}, where the scaling dimensions of the long multiplet are given by
\es{superMean}{
\Delta_{(A,0)_{j,n}}^{\text{GFFT}}=j+2+2n\,,
}
with $n=0, 1, 2, \ldots$.

 \begin{table}[htp]
\begin{center}
\begin{tabular}{|l|c|c|}
\hline
 \multicolumn{1}{|c|}{Type $\cal M$} & Free theory $\lambda_{\cal M}^2$  & generalized free field theory $\lambda_{\cal M}^2$ \\
  \hline
  $(B,2)$ & $\,\;\quad\qquad0$ & $\;\;\quad\qquad32/3\approx10.667$\\
  $(B,+)$ & $\,\;\quad\qquad16$ & $\;\;\quad\qquad16/3\approx5.333$\\
  $(A,2)_1$ & $\,\;\quad\qquad128/21\approx6.095$ & $\;\;\quad\qquad1024 / 105\approx9.752$\\
  $(A,2)_3$ & $\quad\qquad2048/165\approx12.412$ & $\;\;\qquad131072 / 8085\approx16.212$\\
  $(A,2)_5$ & $9273344 / 495495\approx18.715$ & $33554432 / 1486485\approx22.573$\\
  $(A,+)_0$ & $\quad\qquad\qquad32/3\approx10.667$ & $\;\;\quad\qquad\qquad64 / 9\approx7.111$\\
  $(A,+)_2$ & $\qquad20992/1225\approx17.136$ & $\quad\qquad16384 / 1225\approx13.375$\\
  $(A,+)_4$ & $\;\;\quad139264 / 5929\approx23.489$ & $\;\;\quad1048576 / 53361\approx19.651$\\
  $(A,0)_{0,0}$ & $\quad4$ & $\quad\qquad\qquad 32 / 35\approx0.911$\\
  $(A,0)_{2,0}$ & $\quad4$ & $\;\;\quad\qquad 2048 / 693\approx2.955$\\
  $(A,0)_{4,0}$ & $\quad4$ & $ \;\;1048576 / 225225\approx4.656$\\
  \hline
\end{tabular}
\end{center}
\caption{Values of OPE coefficients in the free and generalized free field theory limits for the $(B,2)$ and $(B,+)$ multiplets, the $(A,2)_j$ multiplet for spin $j=1,3,5$, the $(A,+)_j$ multiplet for $j=0,2,4$, and the $(A,0)_{j,n}$ multiplet for $j=0,2,4$ and $n=0$, which is the lowest multiplet with that spin.}\label{Avalues}
\end{table}

\section{Computation of OPE coefficients}
\label{1d}

A crucial input into the numerical bootstrap analysis, which we will use to isolate the ${\cal N} =8$ SCFTs with holographic duals, is that we can compute the squared OPE coefficients $\lambda_\text{Stress}^2$, $\lambda_{(B, +)}^2$, and $\lambda_{(B, 2)}^2$ in all Lagrangian ${\cal N} =8$ SCFTs using supersymmetric localization.  Conceptually, this computation can be split into several parts, each of which we discuss in separate subsections.  In Section~\ref{TOP}, we explain how, in ${\cal N} =4$ SCFTs with flavor symmetries, one can relate the fourth derivative of the mass-deformed $S^3$ partition function with respect to the mass parameter to certain OPE coefficients.  In Section~\ref{APPN8}, we apply this analysis to ${\cal N} = 8$ SCFTs.  In Section~\ref{EXAMPLES}, we use the existing results for the mass-deformed partition function of ABJM and BLG theories in order to extract $\lambda_\text{Stress}^2$, $\lambda_{(B, +)}^2$, and $\lambda_{(B, 2)}^2$ from the results of the previous two sections.

\subsection{Topological sector of ${\cal N} = 4$ SCFTs from mass-deformed $S^3$ partition function}
\label{TOP}

In this subsection, let us discuss some general results for ${\cal N} = 4$ SCFTs.  ${\cal N} = 4$ SCFTs are invariant under the superconfomal algebra $\mathfrak{osp}(4|4)$, which contains, as its bosonic subalgebra, the conformal algebra $\mathfrak{so}(3, 2)$ as well as an $\mathfrak{so}(4)_R$ symmetry which we write as $\mathfrak{su}(2)_H \oplus \mathfrak{su}(2)_C$.  We will use $a, b, c$, etc.~for the $\mathfrak{su}(2)_H$ fundamental indices and $\dot a, \dot b, \dot c$, etc.~for the $\mathfrak{su}(2)_C$ fundamental indices.  These indices are raised and lowered with the anti-symmetric $\epsilon$ symbol, for instance $\phi^a = \epsilon^{ab} \phi_b$ or $\phi_a = \epsilon_{ab} \phi^b$, with $\epsilon^{12} = - \epsilon_{12} = 1$.

It was shown in \cite{Beem:2016cbd,Chester:2014mea} that each 3d ${\cal N} = 4$ SCFT contains 1d topological sectors that capture information on the half-BPS spectrum of the 3d theory.  There are two such sectors whose abstract constructions mirror each other by simply flipping the roles played by $\mathfrak{su}(2)_H$ and $\mathfrak{su}(2)_C$.  For our final goal of computing $\lambda_\text{Stress}^2$, $\lambda_{(B, +)}^2$, and $\lambda_{(B, 2)}^2$ in a given ${\cal N} = 8$ theory, it does not matter which of the two 1d sectors of the 3d ${\cal N} = 4$ SCFT we focus on, so let us focus on the one built from $\mathfrak{su}(2)_C$-invariant operators. (This sector is associated with the Higgs branch in the terminology of \cite{Beem:2016cbd,Chester:2014mea}.)  The construction of the 1d operators that descend from 3d local operators is quite simple and proceeds as follows.   The 3d theory generically has $1/2$-BPS ``Higgs branch'' operators ${\cal O}_{a_1 \ldots a_{2 j_H}}(\vec{x})$ that are Lorentz scalars invariant under $\mathfrak{su}(2)_C$ and that have scaling dimension equal to the $\mathfrak{su}(2)_H$ spin, $\Delta = j_H$.  If the 1d sector lies along the $x_3$ axis parameterized by $\vec{x} = (0, 0, x)$, then the 1d operators are
   \es{1dDef}{
    {\cal O}(x) = {\cal O}_{a_1 \ldots a_{2 j_H}}(0, 0, x) u^{a_1}(x) \cdots u^{a_{2j_H}}(x) \,, \qquad
     u^1(x) = 1\,, \qquad u^2(x) =  \frac{x}{2r}\,.
   }
Here, $r$ is a parameter with dimensions of length that was introduced in order for the expression for ${\cal O}(x)$ to be dimensionally correct.  That the operators \eqref{1dDef} are topological follows abstractly from properties of the superconformal algebra:  these operators are in the cohomology of a nilpotent supercharge with respect to which translations in $x$ are exact.

While \eqref{1dDef} was written for the case of an SCFT defined on $\R^3$, one can perform a similar construction on any conformally flat space.  In particular, using the stereographic projection, the 3d SCFT can also be placed on a round three-sphere of radius $r$ such that the 1d line gets mapped to a great circle parameterized by $\varphi = 2 \arctan \frac{x}{2r}$.  In this case the 1d operators ${\cal O}(\varphi)$ are periodic on this circle if $j_H$ is an integer and anti-periodic if $j_H$ is a half odd-integer.  Defining the 1d theory on a great circle of $S^3$ as opposed to a line in $\R^3$ has the benefit that when the 3d SCFT has a Lagrangian description, then it is possible to perform supersymmetric localization on $S^3$ in order to obtain an explicit Lagrangian description of the 1d sector itself.  In the case where the Lagrangian of the 3d theory involves only hypermultiplets and vector multiplets, the 1d theory Lagrangian was derived in \cite{Dedushenko:2016jxl}.

Regardless of whether the 1d theory has a Lagrangian description or not, let us describe a procedure for calculating certain integrated correlation functions in the 1d theory.  We will be interested in the case where the 1d operator $J(\varphi)$ comes from a 3d operator $J_{ab}(\vec{x})$ with $\Delta = j_H = 1$.  Such a 3d operator is the bottom component of a superconformal multiplet $(J_{ab}, K_{\dot a \dot b}, j_\mu, \chi_{a \dot a})$ that in addition to $J_{ab}$ also contains the following conformal primaries:  a pseudoscalar $K_{\dot a \dot b}$ of scaling dimension $2$, a fermion $\chi_{a \dot a}$ of scaling dimension $3/2$, and a conserved current $j_\mu$. %\footnote{For a free hypermultiplet $(q_a, \tq_a, \psi_{\alpha \dot a}, \tpsi_{\alpha \dot a})$, with $\langle q_a(\vec{x}) \tq_b(0) \rangle = \frac{\epsilon_{ab}}{4 \pi \abs{\vec{x}}}$ and $\langle \psi_{\alpha \dot a }(\vec{x}) \tpsi_{\beta \dot b }(0) \rangle = \frac{\epsilon_{ab}x^\mu \gamma_{\mu \alpha \beta}}{4 \pi \abs{\vec{x}}}$, we have a $U(1)$ subgroup of the flavor symmetry under which the hypermultiplet fields have charges $(1, -1, 1, -1)$.  In this case, $j_\mu = \epsilon^{ab} \left( \tq_a \partial_\mu q_b -  q_a \partial_\mu \tq_b\right) + \epsilon^{\dot a \dot b} \psi_{\alpha a} \gamma_{\mu}^{\alpha \beta} \tpsi_{\beta b}  $, $J_{ab} = \frac 12 \left( q_a \tq_b + q_b \tq_a\right)$ and $K_{\dot a \dot b} = \frac 12 \epsilon^{\alpha \beta} \left( \psi_{\alpha \dot a} \tpsi_{\beta \dot b} + \psi_{\alpha \dot b} \tpsi_{\beta \dot a} \right)$.}  
Thus, this is a conserved flavor current multiplet, and all its operators transform in the adjoint representation of the flavor symmetry.  To exhibit the adjoint indices, we will write $(J_{ab}^A, K_{\dot a \dot b}^A, j_\mu^A, \chi_{a \dot a}^A)$, where $A$ runs from $1$ to the dimension of the flavor symmetry Lie algebra.

Let us choose a basis for this Lie algebra where the two-point function of the current multiplets is diagonal in the adjoint indices:
  \es{Normalizations}{
   \langle j_\mu^A(\vec{x}) j_\nu^B(0) \rangle &= \delta^{AB} \frac{\tau}{64 \pi^2} \left(\partial^2 \delta_{\mu\nu} - \partial_\mu \partial_\nu \right) \frac{1}{\abs{\vec{x}}^2} \,, \\
   \langle J_{ab}^A(\vec{x}) J_{cd}^B(0) \rangle &= - \delta^{AB} \frac{\tau}{64 \pi^2} \frac{1}{\abs{\vec{x}}^2} (\epsilon_{ac} \epsilon_{bd} + \epsilon_{ad} \epsilon_{bc}) \,, \\
   \langle K_{\dot a \dot b}^A(\vec{x}) K_{\dot c\dot d}^B(0) \rangle &= - \delta^{AB} \frac{\tau}{32 \pi^2} \frac{1}{\abs{\vec{x}}^4} (\epsilon_{\dot a \dot c} \epsilon_{\dot b \dot d} + \epsilon_{\dot a \dot d} \epsilon_{\dot b \dot c}) \,.
  }
Let us also normalize the current $j_\mu^A$ canonically, meaning that for an operator ${\cal O}$ transforming in a representation ${\cal R}$ of the flavor symmetry, we have $j_\mu^A(\vec{x}) {\cal O}(0) \sim \frac{x_\mu}{4 \pi \abs{\vec{x}}^3} T^A {\cal O}(0)$, where $T^A$ is the corresponding generator in representation ${\cal R}$.  In particular, we then have $j_\mu^A(\vec{x}) j_\nu^B(0) \sim \frac{x_\mu}{4 \pi \abs{\vec{x}}^3} i f^{ABC} j_\nu^C(0)$, where the structure constants are defined by $[T^A, T^B] = i f^{ABC} T^C$.\footnote{For example, a free hypermultiplet has $\mathfrak{su}(2)$ flavor symmetry and a current multiplet as described.  Indeed, we can write the hyper scalars as $q_{ai}$ and the hyper fermions as $\psi_{\dot a i}$, with the two-point functions $\langle q_{ai}(\vec{x}) q_{bj}(0) \rangle = \frac{\epsilon_{ab} \epsilon_{ij}}{4 \pi \abs{x}}$ and $\langle \psi_{\alpha \dot a i }(\vec{x}) \psi_{\beta \dot b j}(0) \rangle = \frac{\epsilon_{ab} \epsilon_{ij} x^\mu \gamma_{\mu \alpha \beta}}{4 \pi \abs{\vec{x}}}$.  Then $j_\mu^A = \frac 12 \sigma^{Aij} \left[i \epsilon^{ab} q_{ai} \partial_\mu q_{bj}  -\frac 12  \epsilon^{\dot a \dot b} \psi_{\alpha \dot a i} \gamma_{\mu}^{\alpha \beta} \psi_{\beta \dot b j} \right] $, $J_{ab}^A =  \frac{1}{8} \sigma^{Aij} (q_{ai} q_{bj} + q_{bi} q_{aj} )$, and $K_{\dot a \dot b}^A =\frac{i}{8} \sigma^{Aij} (\psi_{\dot ai} \psi_{\dot bj} + \psi_{\dot bi} \psi_{\dot aj} )$.  We have $\tau = 1/2$ for the $\mathfrak{su}(2)$ flavor symmetry of a hyper.\label{FreeHyperFootnote}}

At the linearized level, such a current multiplet couples to a background ${\cal N} = 4$ vector multiplet $(A_\mu^A, \Phi_{\dot a \dot b}^A, D_{ab}^A, \lambda_{a \dot a}^A)$:
  \es{S3dLinGen}{
   \int d^3 x\, \left[A^{\mu A} j_\mu^A + i D^{abA} J_{ab}^A + \Phi^{\dot a \dot b A} K_{\dot a \dot b}^A + \text{(fermions)} \right] \,.
  }
(Quadratic terms in the background vector multiplet are also required in order to preserve gauge invariance and supersymmetry.)

Let us provide a prescription for computing correlation functions of the integrated operator $\int d\varphi\, J^A(\varphi)$.  To obtain this prescription, first place the SCFT on a round $S^3$, then introduce an ${\cal N} = 4$-preserving (adjoint valued) real mass parameter $m = m^A T^A$.  Introducing such a parameter requires the following background vector multiplet fields:
 \es{BackgroundVector}{
  \Phi_{\dot a \dot b }^A = m^A \bar h_{\dot a \dot b} \,, \qquad D_{ab}^A =  -\frac{m^A}{r} h_{ab}  \,, \qquad A_\mu^A = \lambda_{a \dot a}^A = 0 \,.
 }
Here, we follow the conventions of \cite{Dedushenko:2016jxl} for the hypermultiplet and vector multiplet fields and their SUSY variations.  The quantities $h$ and $\bar h$ are constant matrices, normalized such that $h_{ab} h^{ab} = \bar h_{\dot a \dot b} \bar h^{\dot a \dot b} = -2$.  The mass-deformed theory is invariant under the superalgebra $\mathfrak{su}(2|1) \oplus \mathfrak{su}(2|1)$, and the mass parameter $m$ appears as a central charge in this algebra.  Up to linear order in $m$, the mass deformation amounts to adding 
   \es{S3dDeformation}{
     m^A \int d^3 x\, \sqrt{g} \left[-i \frac 1r h^{ab} J_{ab}^A + \bar h^{\dot a \dot b} K_{\dot a \dot b}^A \right] 
   }
to the conformal action on $S^3$.  The $S^3$ partition function $Z(m)$ can be computed using supersymmetric localization \cite{Kapustin:2009kz} even for ${\cal N} = 4$ theories for which the localization to the 1d sector performed in \cite{Dedushenko:2016jxl} does not apply.

However, $Z(m)$  also computes the partition function of the 1d theory deformed by
  \es{S1dDef}{
    -4 \pi r^2 m^A \int_{-\pi}^\pi d\varphi  J^A(\varphi) \,.
   }
Such a statement can be proven explicitly\footnote{Let us consider $N$ hypermultiplets $(q_{ai}, \psi_{\dot a i})$, where $i = 1, \ldots, 2N$, charged under a vector multiplet with gauge group $G$ and generators $t^\alpha$ and flavor symmetry $G_F$ with generators $T^{A}$, respectively.  Both $G$ and $G_F$ are embedded into the flavor symmetry $USp(2N)$ of $N$ ungauged free hypers.  We then have $J_{ab}^A =  \frac{1}{4} T^{Aij} (q_{ai} q_{bj} + q_{bi} q_{aj} )$, where $T^{Aij}$ is a symmetric matrix in the $ij$ indices.  Consequently, using \eqref{1dDef}, we have $J^A = J_{ab}^A u^a u^b= \frac{1}{4} T^{Aij} Q_{i} Q_{j}$, with $Q_i = q_{ai} u^a$.  In \cite{Dedushenko:2016jxl}, it was shown that the partition function of the 1d topological theory, defined on a circle, is described by the partition function
 \es{Z}{
  Z = \int_{\text{Cartan of $\mathfrak{g}$}} d\sigma\, \det{}_\text{adj} (2 \sinh (\pi \sigma)) \int DQ\, e^{- 2\pi r \int d\varphi\, \left(\Omega^{ij} Q_i \partial_\vphi Q_j 
   -  \sigma^{\alpha} t^{\alpha ij} Q_i Q_j  - r m^A T^{Aij} Q_i Q_j \right) }\,.
 }
So deforming the 3d theory by a mass parameter $m$ is equivalent to deforming the 1d theory by \eqref{S1dDef}.  For a single free hyper, we have $N=1$, $\Omega^{ij} = \epsilon^{ij}$ and $T^{Aij} = \frac 12 \sigma^{Aij}$ for the $SU(2)$ flavor symmetry---see Footnote~\ref{FreeHyperFootnote}.  
} in the case where the results of \cite{Dedushenko:2016jxl} apply, but it should hold more generally.  This statement should simply follow from the supersymmetric Ward identities, as was shown in similar 4d examples in~\cite{Gerchkovitz:2016gxx,Gomis:2014woa};  it would be nice to investigate this more precisely in the future.  In other words, we claim that the supersymmetric Ward identity must imply that the expressions \eqref{S3dDeformation} and \eqref{S1dDef} are equal up to $Q$-exact terms.  

Consequently, we have that
 \es{IntegratedCorrelator}{
  \left\langle  \int d\varphi J^{A_1}(\varphi) \cdots \int d\varphi J^{A_n}(\varphi) \right \rangle = \frac{1}{(4 \pi r^2)^n} \frac{1}{Z} \frac{d^n Z}{dm^{A_1} dm^{A_2} \cdots dm^{A_n}} \bigg|_{m=0} \,.
 }
This is the main result of this subsection.

As a particular case, we can consider $n=2$.  From \eqref{Normalizations}, we see that on a line in $\R^3$, we have $\langle J^A(x) J^B(0) \rangle = -\frac{\tau}{128 \pi^2 r^2} \delta^{AB}$, and so 
 \es{IntegratedCorrelator2}{
  \left \langle  \int d\varphi J^A(\varphi)   \int d\varphi J^B(\varphi)  \right \rangle =  -\frac{\tau}{32 r^2} \delta^{AB} \,.
 }
Comparing to \eqref{IntegratedCorrelator}, we deduce
 \es{Gottau}{
  \tau = -\frac{2}{\pi^2} \frac{1}{Z} \frac{d^2 Z}{d(r m^A)^2} \bigg|_{m=0}  =  \frac{2}{\pi^2} \frac{d^2F_{S^3}}{d(r m^A)^2} \bigg|_{m=0} 
 }
(with no summation over $A$), where we defined the $S^3$ free energy $F_{S^3} = - \log Z$.  This formula agrees with the result of \cite{Closset:2012vg}.  (For an ${\cal N} = 4$ mass-deformed SCFT on $S^3$, the free energy is real, so one does not have to take its real part as in \cite{Closset:2012vg}.)

\subsection{Application to ${\cal N} = 8$ SCFTs}
\label{APPN8}

In order to apply the above results to ${\cal N} = 8$ SCFTs, one would first have to go through the exercise of decomposing the various representations of the ${\cal N} = 8$ superconformal algebra into representations of the ${\cal N} = 4$ superconformal algebra in order to establish which of the ${\cal N} = 8$ irreps contain Higgs branch scalar operators with $\Delta = j_H$ and $j_C = 0$.  In the notation introduced in Table~\ref{Multiplets}, it can be checked that these irreps are of $(B, 2)$, $(B, 3)$, $(B, +)$, and $(B, -)$ type.  So these are the ${\cal N} = 8$ multiplets that are captured by the 1d topological sector discussed above.  In performing the decomposition from ${\cal N} = 8$ to ${\cal N} = 4$, one should also keep track of an $\mathfrak{su}(2)_F \oplus \mathfrak{su}(2)_{F'}$ flavor symmetry that commutes with $\mathfrak{so}(4)_R$ inside $\mathfrak{so}(8)$.

As mentioned in Section~\ref{4point}, we are interested in analyzing the 4-point function of the ${\bf 35}_c = [0020]$ scalar in the same ${\cal N} = 8$ superconformal multiplet as the stress tensor, so we should only focus on the stress tensor multiplet ($(B, +)_{[0020]}$) as well as the multiplets $(B, +)_{[0040]}$ and $(B, 2)_{[0200]}$ (for short referred to as $(B, +)$, and $(B, 2)$ above) that appear in the OPE of two stress tensor multiplets.  These multiplets contain the following Higgs branch operators (HBOs)
 \es{HiggsFromN8}{
  \text{Stress} &\supset \text{HBOs with $\Delta = j_H = 1$ in $({\bf 3}, {\bf 1})$ of flavor $\mathfrak{su}(2)_F \oplus \mathfrak{su}(2)_{F'}$} \,, \\
  (B, +) &\supset \text{HBOs with $\Delta = j_H = 2$ in $({\bf 5}, {\bf 1})$ of flavor $\mathfrak{su}(2)_F \oplus \mathfrak{su}(2)_{F'}$} \,, \\
  (B, 2) &\supset \text{HBOs with $\Delta = j_H = 2$ in $({\bf 1}, {\bf 1})$ of flavor $\mathfrak{su}(2)_F \oplus \mathfrak{su}(2)_{F'}$}  \,.
 }

Thus, from the ${\cal N} = 4$ perspective, each local ${\cal N} = 8$ SCFT contains a conserved current multiplet $(J_{ab}^A, K_{\dot a \dot b}^A, j_\mu^A, \chi_{a \dot a}^A)$, which transforms in the adjoint of an $\mathfrak{su}(2)_F$ flavor symmetry ($A = 1, 2,3$ in this case).  This multiplet is embedded in the ${\cal N} = 8$ stress tensor multiplet, with $\mathfrak{su}(2)_F$ embedded into $\mathfrak{so}(8)_R$.  Consequently, the coefficient $\tau$ appearing in the two-point function \eqref{Normalizations} of the canonically normalized currents must be proportional to coefficient $c_T$ appearing in the two-point function \eqref{CanStress} of the canonically normalized stress tensor.  In a free ${\cal N} = 8$ theory we have\footnote{In ${\cal N} = 4$ notation, an ${\cal N} = 8$ free theory is a product between a theory of a free hypermultiplet and a free twisted hypermultiplet.  The $\mathfrak{su}(2)_F$ acts on the hypermultiplet only, so $\tau$ has the same value as in the free hypermultiplet theory, namely $\tau = 1/2$, as explained in Footnote~\ref{FreeHyperFootnote}.}  $c_T = 16$ and $\tau = 1/2$, so
 \es{taucTRelation}{
  \tau = \frac{c_T}{32} \,.
 }

The precise projection to 1d was performed in \cite{Chester:2014mea}.  Converting to the notation here,\footnote{Up to an overall constant, Eq.~(3.31) of \cite{Chester:2014mea} gives the 4-point function of $J^A \sigma^A_{ij} \bar y^i \bar y^j$, where $\bar y^i$ are auxiliary polarization variables, and $\sigma^A_{ij} \equiv (-i \sigma_2 \tau^A)_{ij}$, where $\tau^A$ are the standard Pauli matrices.} we have
\es{JFourPoint}{
\langle J^A(\varphi_1)J^B(\varphi_2)J^C(\varphi_3)J^D(\varphi_4)\rangle&= 
 \frac{\tau^2}{(128 \pi^2 r^2)^2}\Biggl[ \left( 1+\frac{1}{16}\lambda^2_{(B,2)} \right) \delta^{AB} \delta^{CD} \\
&{}+\frac 14 \sgn(\varphi_{12}\varphi_{13}\varphi_{24}\varphi_{34})\lambda^2_\text{Stress} (\delta^{AC} \delta^{BD} - \delta^{AD} \delta^{BC})\\
&{}+\frac{3}{16}\lambda^2_{(B,+)}\left(\delta^{AC} \delta^{BD} + \delta^{AD} \delta^{BC} - \frac{2}{3} \delta^{AB} \delta^{CD} \right) \Biggr] \,.
}
The crossing symmetry of this four-point function implies
\es{crossConstraints}{
4{ \lambda}^2_\text{Stress} - 5 { \lambda}^2_{(B,+)} +{ \lambda}^2_{(B,2)} + 16 = 0\,.
}

Let us now use the results of the previous section in order to extract the OPE coefficients $\lambda_\text{Stress}^2$, $\lambda_{(B, +)}^2$, and $\lambda_{(B, 2)}^2$.  Eq~\eqref{taucTRelation} gives a way to compute $\lambda_\text{Stress}^2$.  From $\lambda_\text{Stress}^2 = 256/c_T$ and \eqref{taucTRelation} we obtain $\lambda_\text{Stress}^2 = 8/\tau$, and from \eqref{Gottau} we further obtain
 \es{GotcT}{
  \lambda_\text{Stress}^2  =  -\frac{4 \pi^2}{\frac{1}{Z} \frac{d^2Z}{d(r m^A)^2} \bigg|_{m=0}} \,, \qquad
   c_T = -\frac{64}{\pi^2} \frac{1}{Z} \frac{d^2Z}{d(r m^A)^2} \bigg|_{m=0} \,.
 }
The other OPE coefficients can be calculated by specializing the four-point function \eqref{JFourPoint} to $A = B = C = D$ and integrating over $\vphi$:
 \es{IntFourPoint}{
  \left\langle \left( d\vphi \int  J^A(\vphi) \right)^4 \right\rangle
   =  \frac{9}{5} \frac{\tau^2}{(32 r^2)^2}\Biggl[ 1+\frac{1}{16}\lambda^2_{(B,2)} +\frac{\lambda_\text{Stress}^2  }{9} \Biggr] \,.
 }
Comparing with \eqref{IntegratedCorrelator}, we obtain
 \es{GotLambdaB2}{
    \lambda^2_{(B,2)}   = 16 \left[ -1 + \frac{4 \pi^2  }{9}  \frac{1}{ \frac{1}{Z} \frac{d^2Z}{d(r m^A)^2}  }+ \frac{5}{9} \frac{\frac{1}{Z} \frac{d^4 Z}{d (r m^{A})^4}  }{\left(\frac{1}{Z}  \frac{d^2Z}{d(r m^A)^2}  \right)^2} \right] \Bigg|_{m=0}\,.
 }
From \eqref{crossConstraints} we can then also obtain $\lambda_{(B, +)}^2$:
 \es{GotLambdaBp}{
  \lambda_{(B, +)}^2 = \frac{16}{9} \left[  -   \frac{\pi^2}{ \frac{1}{Z} \frac{d^2Z}{d(r m^A)^2}  } + \frac{\frac{1}{Z} \frac{d^4 Z}{d (r m^{A})^4}  }{\left(\frac{1}{Z}  \frac{d^2Z}{d(r m^A)^2}  \right)^2} \right] \Bigg|_{m=0}  \,.
 }

\subsection{OPE coefficients in BLG and ABJ(M) theory}
\label{EXAMPLES}

Let us now apply the formulas \eqref{GotcT}, \eqref{GotLambdaB2}, and \eqref{GotLambdaBp} to the specific examples of ABJ(M) and BLG theories.  For simplicity, let us turn on a mass parameter $m$ through the Cartan of $\mathfrak{su}(2)_F$, thus dropping the superscript $A$ from \eqref{GotcT}--\eqref{GotLambdaBp}.  We also set $r = 1$ for simplicity.

The mass-deformed partition function of the $U(N)_k \times U(M)_{-k}$ ABJ(M) theory takes the form
 \es{MassDeformedABJM}{
  Z_\text{ABJ(M)} (m) &= \frac{1}{N! M!} \int d^N\lambda d^M\mu e^{i\pi k\left[\sum_i\lambda^2_i-\sum_j\mu_j^2\right]} \\
  &{}\times \frac{\prod_{i < j}\left( 4 \sinh^2[\pi(\lambda_i-\lambda_j)] \right) \prod_{i < j}\left( 4\sinh^2[\pi(\mu_i-\mu_j)] \right) }{\prod_{i,j}\left( 4 \cosh[\pi(\lambda_i-\mu_j+m/2)]  \cosh[\pi(\mu_i-\lambda_j)] \right) }\,,
 }
For BLG theory, we take $N=M=2$ in the formula above and insert $\delta(\lambda_1 + \lambda_2) \delta(\mu_1 + \mu_2)$ in the integrand, thus obtaining
 \es{MassDeformedBLG}{
  Z_\text{BLG} (m) = \frac{1}{32} \int d\lambda d\mu e^{2 \pi i k\left[\lambda^2-\mu^2\right]}\frac{\sinh^2(2\pi \lambda)\sinh^2(2\pi \mu)}{\prod_{i,j}\cosh[\pi(\lambda_i-\mu_j+m/2)]\cosh[\pi(\mu_i-\lambda_j)]}\,,
 }
where now $\lambda_i = (\lambda, -\lambda)$ and $\mu_i = (\mu, -\mu)$.

\subsubsection{BLG theory}

For BLG theory, one can use the identity
 \es{Identity}{
   \frac{\sinh(2\pi \lambda)\sinh(2\pi \mu)}{4 \prod_{i,j}\cosh[\pi(\lambda_i-\mu_j+m/2)]}
   = \det \begin{pmatrix}
    \frac{1}{2 \cosh [\pi (\lambda - \mu + m)]} & \frac{1}{2 \cosh [\pi (\lambda + \mu + m)]} \\
    \frac{1}{2 \cosh [\pi (-\lambda - \mu + m)]} & \frac{1}{2 \cosh [\pi (-\lambda + \mu + m)]} 
   \end{pmatrix}
 }
and change variables to $x = (\lambda + \mu)/2$ and $y = (\lambda - \mu)/2$, to show that
 \es{ZBLGm}{
   Z_\text{BLG}(m) = \frac{k}{32} \int dx \, \frac{x}{\sinh(\pi k x)} \left[\sech^2 \left( \frac{m \pi}{2} \right) - \sech \left( 
    \frac{m \pi}{2} -x \right)  \sech \left( 
    \frac{m \pi}{2} +x \right)  \right] \,.
 } 
One can then plug this expression into \eqref{GotcT}--\eqref{GotLambdaBp}, which gives 
 \es{lambdasBLG}{
  \lambda_\text{Stress}^2 &= \frac{8 I_{2, k}}{2 I_{2, k} - I_{4, k}} \,, \qquad c_T= 32 \left(2-\frac{I_{4, k}}{I_{2, k}}\right) \,, \\
\lambda^2_{(B,2)}&=\frac{16\left( 6I_{2, k}^2-3I_{4, k}^2-12I_{2, k}I_{4, k}+10I_{2, k} I_{6, k}\right)}{3 (I_{4, k}-2I_{2, k})^2} \,,  \\
  \lambda_{(B, +)}^2 &= \frac{32 I_{2, k} \left(3 I_{2, k} - 3 I_{4, k} + I_{6, k} \right) }{3 (I_{4, k}-2I_{2, k})^2 } \,,
 }
where we defined the integral
 \es{IDef}{
  I_{n, k} \equiv \int_{-\infty}^\infty dx\, \frac{x}{\sinh(\pi k x)} \tanh^n (\pi x) \,.
 }
This integral can be evaluated explicitly using contour integration. We give the expressions for $n=2, 4, 6$ in the Appendix.

\subsubsection{ABJ(M) theory}

For ABJ(M) theory, one can use \eqref{MassDeformedABJM} and \eqref{GotcT}--\eqref{GotLambdaBp} to evaluate $\lambda_\text{Stress}^2$, $\lambda_{(B, +)}^2$, and $\lambda_{(B, 2)}^2$. The number of integrals increases with $N$, however, and unlike the BLG case where analytical formulas were possible for the entire family of theories, in the ABJ(M) case we can perform these integrals analytically only for small values of $N$---See Table~\ref{valList}.

\begin{table}[!h]
\begin{center}
\begin{tabular}{c||c|c|c}
 & $\frac{\lambda_\text{Stress}^2}{16}=\frac{16}{c_T}$ & $\lambda_{(B,+)}^2$ & $\lambda_{(B,2)}^2$\\
 \hline \hline
 ABJM$_{1,k}$   &   $1$  & $16$  &   $0$ \\
  ABJM$^\text{int}_{2,1}\cong \text{ABJ}
  _1$  & $\frac34\approx 0.75$   & $\frac{64}{5}\approx12.8$  &   $0$ \\
  BLG$_1\cong\text{ABJM}_{2,1}$   &   $\frac37\approx0.429$  & $\frac{384}{245} \approx 9.731$  &   $\frac{256}{49} \approx 5.224$ \\
 $ \text{ABJM}_{2,2}\cong\text{BLG}_2\cong\text{ABJ}_{1}^2$& $\frac38\approx 0.375$   & $\frac{136}{15}\approx9.067$  &   $\frac{16}{3}\approx5.333$ \\
 {BLG}$_3\cong$ABJM$^\text{int}_{3,1}$  &  $\frac{\pi-3}{10\pi-31}\approx0.340$  & $\frac{16(\pi-3)(840\pi-2629)}{15(10\pi-31)^2} \approx8.676$ & $ \frac{62208+16\pi(420\pi-2557)}{3(10\pi-31)^2}\approx 5.593$ \\
   BLG$_4\cong$ABJ$_2$ & $\frac{3\pi^2-24}{18\pi^2-160}\approx0.318$&  $\frac{32(\pi^2-8)(315\pi^2-2944)}{15(80-9\pi^2)^2}\approx8.444$ &  $\frac{16(16384-3872\pi^2+225\pi^4)}{3(80-9\pi^2)^2}\approx5.883$ \\
   BLG$_{5}$ & $0.302$&  8.300 &  6.156 \\
 $\vdots$ & $\vdots$  & $\vdots$ & $\vdots$ \\
  BLG$_\infty$   &    $\frac14\approx0.25$     & $8$  &   $8$ \\
   ABJ(M)$_\infty$   &    $0$     & $\frac{16}{3}\approx5.333$  &   $\frac{32}{3}\approx10.666$ \\
    \hline
  \end{tabular}
\caption{OPE coefficients of $\frac12$ and $\frac14$ BPS operators that appear in $\cO_\text{Stress}\times\cO_\text{Stress}$ for $\mathcal{N}=8$ theories. ``$\cong$'' denotes that theories have the same stress tensor four-point function. \label{valList}}
\end{center}
\end{table}

One can also perform a $1/N$ expansion, where $M=N$ or $M=N+1$. There are several approaches for developing a $1/N$ expansion:  one can either work more generally in the 't Hooft limit where $N$ is taken to be large while $N/k$ is held fixed \cite{Drukker:2010nc}, and then take $N/k$ large;  or one can work at fixed $k$ while taking $N$ large \cite{Herzog:2010hf, Marino:2011eh}.  We will follow the approach originating in \cite{Marino:2011eh}, where for $m=0$ it was noticed in \cite{Marino:2011eh} that the $S^3$ partition function for ABJM theory can be rewritten as a partition function of a non-interacting Fermi gas of $N$ particles with kinetic energy $T(p) = \log \cosh (\pi p)$ and potential energy $U(q) = \log \cosh (\pi q)$.  Phase space quantization and statistical physics techniques allow one to calculate the $S^3$ partition function to all orders in $1/N$, and this expansion resums into an Airy function.  The $S^3$ partition function in the presence of a mass deformation was computed using the same method in~\cite{Nosaka:2015iiw}. 

Up to non-perturbative corrections in $1/N$ and an overall $m$-independent prefactor, the result of \cite{Nosaka:2015iiw} gives\footnote{The result of~\cite{Nosaka:2015iiw} is only for $N=M$, but here we generalize it to $N \neq M$ using the results of \cite{Honda:2014npa}.  }${}^{,}$\footnote{In the notation of  \cite{Nosaka:2015iiw}, we can take $\zeta_\lambda = i m$ and $\zeta_\mu = 0$.} 
\es{largeNZ}{
Z(m)& \approx e^A C^{-\frac13}\text{Ai}\left[C^{-\frac13}(N-B)\right]\,,\\
C&=\frac{2}{\pi^2k(1+m^2)}\,,\qquad B=\frac{\pi^2 C}{3}- \frac{2 + m^2}{6k(1+m^2)}-\frac{k}{12}+\frac{k}{2}\left(\frac12-\frac{M-N}{k}\right)^2\,,\\
A&=\frac14\left({\cal A}[k(1+im)]+{\cal A}[k(1-im)]+2 {\cal A} [k]\right)\,,
}
where $k>0$, $M\geq N$, and the function ${\cal A}$ is given by
\es{constantMap}{
{\cal A}(k)=\frac{2\zeta(3)}{\pi^2k}\left(1-\frac{k^3}{16}\right)+\frac{k^2}{\pi^2}\int_0^\infty dx\frac{x}{e^{kx}-1}\log\left(1-e^{-2x}\right)\,.
}
In order to plug this expression into \eqref{GotcT}--\eqref{GotLambdaBp}, one needs the following derivatives of ${\cal A}$:
\es{Aprime}{
{\cal A}''(1)&=\frac16+\frac{\pi^2}{32}\,,\qquad\qquad \,\, \,\,{\cal A}''(2)=\frac{1}{24}\,,\\
{\cal A}''''(1)&=1+\frac{4\pi^2}{5}-\frac{\pi^4}{32}\,,\qquad {\cal A}''''(2)=\frac{1}{16}+\frac{\pi^2}{80}\,.
} 
Using \eqref{GotcT}, we then find
\es{cTABJM}{
c_T^{\text{ABJM}_{N,1}}&=1-\frac{112}{3\pi^2}-\frac{8(9+8N)\text{Ai}'\left[\left(N-3/8\right)({\pi^2}/{2})^{1/3}\right]}{3(\pi^2/2)^{2/3} \text{Ai}\left[\left(N-{3}/{8}\right)({\pi^2}/{2})^{1/3}\right] }\,,\\
c_T^{\text{ABJM}_{N,2}}&=-\frac{112}{3\pi^2}-\frac{64(1+2N)\pi^{2/3}\text{Ai}'\left[\left(N-1/4\right){\pi}^{2/3}\right]}{3\pi^2 \text{Ai}\left[\left(N-{1}/{4}\right){\pi}^{2/3}\right] }\,,\\
c_T^{\text{ABJ}_{N}}&=-\frac{112}{3\pi^2}-\frac{32(3+4N)\pi^{2/3}\text{Ai}'\left[N\pi^{2/3}\right]}{3\pi^{2} \text{Ai}\left[N{\pi}^{2/3}\right] }\,,
}
which are expressions valid to all orders in $1/N$.   The analogous expressions for $\lambda^2_{(B,2)}$ are rather complicated, so we delegate them to Appendix~\ref{1dmore}.  The formulas for $\lambda^2_{(B,+)}$ can then be determined from \eqref{crossConstraints}.

A comment is in order for the ABJM$_{N, 1}$ theory.  This theory is a direct product between a free sector, identified with ABJM$_{1, 1}$, and an interacting sector.  Since the free sector has $c_T = 16$, we have that the interacting sector of the ABJM$_{N, 1}$ theory has
 \es{cTInt}{  
  c_{T, \text{int}}^{\text{ABJM}_{N,1}}=c_T^{\text{ABJM}_{N,1}}-16 \,.
 }
Extracting the value of $\lambda^2_{(B,2)}$ of just the interacting sector knowing $\lambda^2_{(B,2)}$ for the full theory requires more thought.  In the free theory one has $\lambda^2_{(B,2)} = 0$, as such a multiplet does not exist as can be checked explicitly by decomposing the 4-point function of ${\cal O}_\text{Stress}$ in superconformal blocks \cite{Chester:2014fya}.  Using this fact, and the general formulas for how squared OPE coefficients combine when taking product CFTs \cite{Chester:2014mea}, one has
\es{B2int}{
\left(\lambda^{\text{ABJM}_{N,1}}_{(B,2)}\right)^2=\frac{2c_{T,\text{free}}^{\text{ABJM}_{N,1}}c_{T,\text{int}}^{\text{ABJM}_{N,1}}\lambda_{(B,2),\text{GFFT}}^2+\left(c_{T,\text{int}}^{\text{ABJM}_{N,1}}\right)^2\left(\lambda^{\text{ABJM}_{N,1}}_{(B,2),\text{int}}\right)^2 }{\left(c_{T}^{\text{ABJM}_{N,1}}\right)^2}\,.
}
where $\lambda_{(B,2),\text{GFFT}}^2 = 32/3$ is the generalized free theory value of the $\lambda^2_{(B,2)}$ OPE coefficient as given in \eqref{Avalues}.

In Table \ref{compTab} we compare $\lambda^2_{(B,2)}$, $\lambda^2_{(B,+)}$ and $c_T$ to the exact values for $N=2$ and find excellent agreement.

\begin{table}[!h]
\begin{center}
\begin{tabular}{c||c|c|c}
 & $\frac{\lambda_\text{Stress}^2}{16}=\frac{16}{c_T}$& $\lambda_{(B,+)}^2$ & $\lambda_{(B,2)}^2$\\
 \hline \hline
 large $N$ ABJM$_{2,1}^\text{int}$   & $0.7500$  & $ 12.7982$  &   $-0.0100$ \\
exact ABJM$_{2,1}^\text{int}$   &   $\frac34\approx0.75$   & $\frac{64}{5} \approx 12.800$ & $0$ \\
\hline
 large $N$ ABJM$_{2,2}$  & $0.3759 $   & $9.0513$  &   $ 5.1995$ \\
 exact ABJM$_{2,2}$   & $\frac38\approx0.375$  & $\frac{136}{15}\approx 9.0667$  &   $\frac{16}{3}\approx 5.3333$ \\
 \hline
 large $N$ ABJ$_{2}$  & $0.3173 $   & $8.4533$  &   $ 5.9618$ \\
 exact ABJ$_{2}$   & $\frac{3\pi^2-24}{18\pi^2-160}\approx0.3177$ & $\frac{32(\pi^2-8)(315\pi^2-2944)}{15(80-9\pi^2)^2}\approx 8.4436$  &   $\frac{16\left(16384-3872\pi^2+225\pi^4\right)}{3(80-9\pi^2)^2}\approx 5.8831$ \\
  \hline
\end{tabular}
\caption{Comparison of the large $N$ formulae to the exact values for $N=2$ for OPE coefficients of $\frac12$ and $\frac14$ BPS operators that appear in $\cO_\text{Stress}\times\cO_\text{Stress}$ for the interacting sector of ABJ(M).  The excellent agreement shows that the asymptotic formulas are reliable for all $N \geq 2$.\label{compTab}}
\end{center}
\end{table}

We can also expand $\lambda^2_{(B,2)}$ and $\lambda^2_{(B,+)}$ directly in terms of $c_T$, by comparing the large $N$ expansions. When expanding the Airy functions in \eqref{cTABJM} and \eqref{B2ABJM} for large $N$, one should be careful to expand in terms of the entire argument of the Airy function, and not just $N$, which is how these functions were originally defined in \cite{Marino:2011eh}. We find that the large $c_T$ expansion is the same for ABJ$_N$ and ABJM$_{N,2}$, while the expansions for ABJM$_{N,k}$ are
 \es{cTexp}{
 \left( \lambda^{\text{ABJM}_{N,k}}_{(B,2)}\right)^2&=\frac{32}{3}-\frac{1024(4\pi^2-15)}{9\pi^2}\frac{1}{c_T^{\text{ABJM}_{N,k}}}+\frac{40960}{\pi^{\frac83}}\left(\frac{2}{9k^2}\right)^{\frac13}\frac{1}{\left(c_T^{\text{ABJM}_{N,k}}\right)^{\frac53}}+O\left({\left(c_T^{\text{ABJM}_{N,k}}\right)^{-2}}\right)\,,\\
  \left(  \lambda^{\text{ABJM}_{N,k}}_{(B,+)}\right)^2&=\frac{16}{3}+\frac{1024(\pi^2+3)}{9\pi^2}\frac{1}{c_T^{\text{ABJM}_{N,k}}}+\frac{8192}{\pi^{\frac83}}\left(\frac{2}{9k^2}\right)^{\frac13}\frac{1}{\left(c_T^{\text{ABJM}_{N,k}}\right)^{\frac53}}+O\left({\left(c_T^{\text{ABJM}_{N,k}}\right)^{-2}}\right)\,,
 }
where note that the leading order correction is independent of $k$.

\section{Bootstrap bound saturation}
\label{min}

\subsection{Numerical bootstrap setup}
\label{bootup}

We will now briefly review how the numerical bootstrap can be applied to the stress-tensor multiplet four-point function in $\mathcal{N}=8$ theories. For further details, see \cite{Chester:2014fya}. Invariance of the four-point function \eqref{FourPointO} under the exchange $(x_1,Y_1)\leftrightarrow(x_3,Y_3)$ implies the crossing equation of the form
 \es{crossingEq}{
 \sum_{{\cal M}\, \in\, \mathfrak{osp}(8|4) \text{ multiplets}} \lambda_{\cal M}^2\,  \vec{d}_{{\cal M}} = 0 \,,
 }
where ${\cal M}$ ranges over all the superconformal multiplets listed in Table~\ref{opemult}, $\vec{d}_{{\cal M}} $ are functions of superconformal blocks, and $\lambda_{\cal M}^2$ are squares of OPE coefficients that must be positive by unitarity.  As in \cite{Chester:2014fya}, we normalize the OPE coefficient of the identity multiplet to $\lambda_{\text{Id}} = 1$, and parameterize our theories by the value of $\lambda_\text{Stress}$, which is related to $c_T$ through \eqref{cTRel}.

To find upper/lower bounds on a given OPE coefficient of a multiplet ${\cal M}'$ that appears in the ${\cal O}_{\text{Stress}} \times {\cal O}_{\text{Stress}}$ OPE, then if $\mathcal{M}'$ is not a long multiplet we consider linear functionals $\alpha$ satisfying
 \es{CondOPE}{
  &\alpha(\vec{d}_{\cal M'}) = s \,, \;\quad\qquad  \text{$s=1$ for upper bounds, $s=-1$ for lower bounds} \,,  \\
  &\alpha(\vec{d}_{\cal M}) \geq 0 \,, \;\;\quad\qquad \text{for all short and semi-short ${\cal M} \notin \{ \text{Id}, \text{Stress}, \cal M' \}$} \,, \\
  &\alpha(\vec{d}_{(A,0)_{j,0}}) \geq 0 \,, \qquad \text{for all $j$ with $\Delta_{(A,0)_{j,0}} \geq j+1$} \,.\\
 }
 If $\mathcal{M}'$ is a long multiplet $(A,0)_{j',n'}$, then we consider linear functionals $\alpha$ satisfying
  \es{CondOPE2}{
  &\alpha(\vec{d}_{\cal M'}) = s \,, \;\quad\qquad\;  \quad\text{$s=1$ for upper bounds, $s=-1$ for lower bounds} \,,  \\
  &\alpha(\vec{d}_{\cal M}) \geq 0 \,, \;\qquad\qquad\;\; \text{for all short and semi-short ${\cal M} \notin \{ \text{Id}, \text{Stress} \}$} \,, \\
 &\alpha(\vec{d}_{(A,0)_{j,0}}) \geq 0 \,,\;\quad\qquad \text{for all $j\neq j'$ with $\Delta_{(A,0)_{j,0}} \geq j+1$} \,,\\
  &\alpha(\vec{d}_{(A,0)_{j',n}}) \geq 0 \,,\; \qquad \;\;\text{for all $n<n'$, and fixed $\Delta_{(A,0)_{j',n}}$} \,,\\
   &\alpha(\vec{d}_{(A,0)_{j',n'+1}}) \geq 0 \,,\qquad \text{with $\Delta_{(A,0)_{j',n'+1}} > \Delta_{(A,0)_{j',n'}}$} \,.\\
 }
 In either case, if such a functional $\alpha$ exists, then this $\alpha$ applied to \eqref{crossingEq} along with the positivity of all $\lambda_{\cal M}^2$ except, possibly, for that of $\lambda_{{\cal M}'}^2$ implies that
 \es{UpperOPE}{
  &\text{if $s=1$, then}\qquad \lambda_{{\cal M}'}^2 \leq - \alpha (\vec{d}_\text{Id})  - \lambda^2_{\text{Stress}} \alpha( \vec{d}_\text{Stress} ) \,,\\
    &\text{if $s=-1$, then}\qquad \lambda_{{\cal M}'}^2 \geq  \alpha (\vec{d}_\text{Id})  + \lambda^2_{\text{Stress}} \alpha( \vec{d}_\text{Stress} ) \,.\\
 }
 Note that the final condition $\Delta_{(A,0)_{j',n'+1}} > \Delta_{(A,0)_{j',n'}}$ when ${\cal M}'$ is a long multiplet $(A,0)_{j',n'}$ is so that ${\cal M}'$ is isolated from the continuum of possible long multiplets. To obtain the most stringent upper/lower bound on $\lambda_{{\cal M}'}^2$, one should then minimize/maximize the RHS of \eqref{UpperOPE} under the constraints \eqref{CondOPE}. 

The numerical implementation of the minimization/maximization problem described above requires two truncations: one in the number of derivatives used to construct $\alpha$ and one in the range of multiplets ${\cal M}$ that we consider. We used the same parameters as in \cite{Simmons-Duffin:2016wlq}, namely spins in $\{0,\dots,64\}\cup\{67,68,71,72,75,76,79,80,83,84,87,88\}$ and derivatives parameter $\Lambda=43$. The truncated minimization/maximization problem can now be rephrased as a semidefinite programing problem using the method developed in \cite{Poland:2011ey}. This problem can be solved efficiently using {\tt SDPB} \cite{Simmons-Duffin:2015qma}.

\subsection{Bounds on OPE coefficients}

Let us now compare the analytical values of the OPE coefficients $\lambda^2_{(B,2)}$, $\lambda^2_{(B,+)}$, and $\lambda^2_\text{Stress}=256/c_T$ found in Section~\ref{4point} to the numerical bootstrap bounds obtained using the method outlined in the previous subsection \cite{Chester:2014fya}.  As noted in \cite{Chester:2014fya}, the numerical bounds on these OPE coefficients exactly satisfy the constraint \eqref{crossConstraints}, so it suffices to discuss the bounds on just two of them, which for simplicity we choose to be $\lambda^2_{(B,2)}$ and $\lambda^2_\text{Stress}$. The main lesson from this comparison will be that $\lambda^2_{(B,2)}$ saturates the lower bounds for all $\mathcal{N}=8$ theories with holographic duals at large $c_T$, so we can use the extremal functional method to read off the spectrum of all operators in the OPE $\cO_\text{Stress}\times \cO_\text{Stress}$ in this regime. At smaller values of $c_T$, we expect that one of these holographic theories saturates the bounds, so the results for the extremal functional hold for that theory.

\begin{figure}[t!]
\begin{center}
   \includegraphics[width=0.85\textwidth]{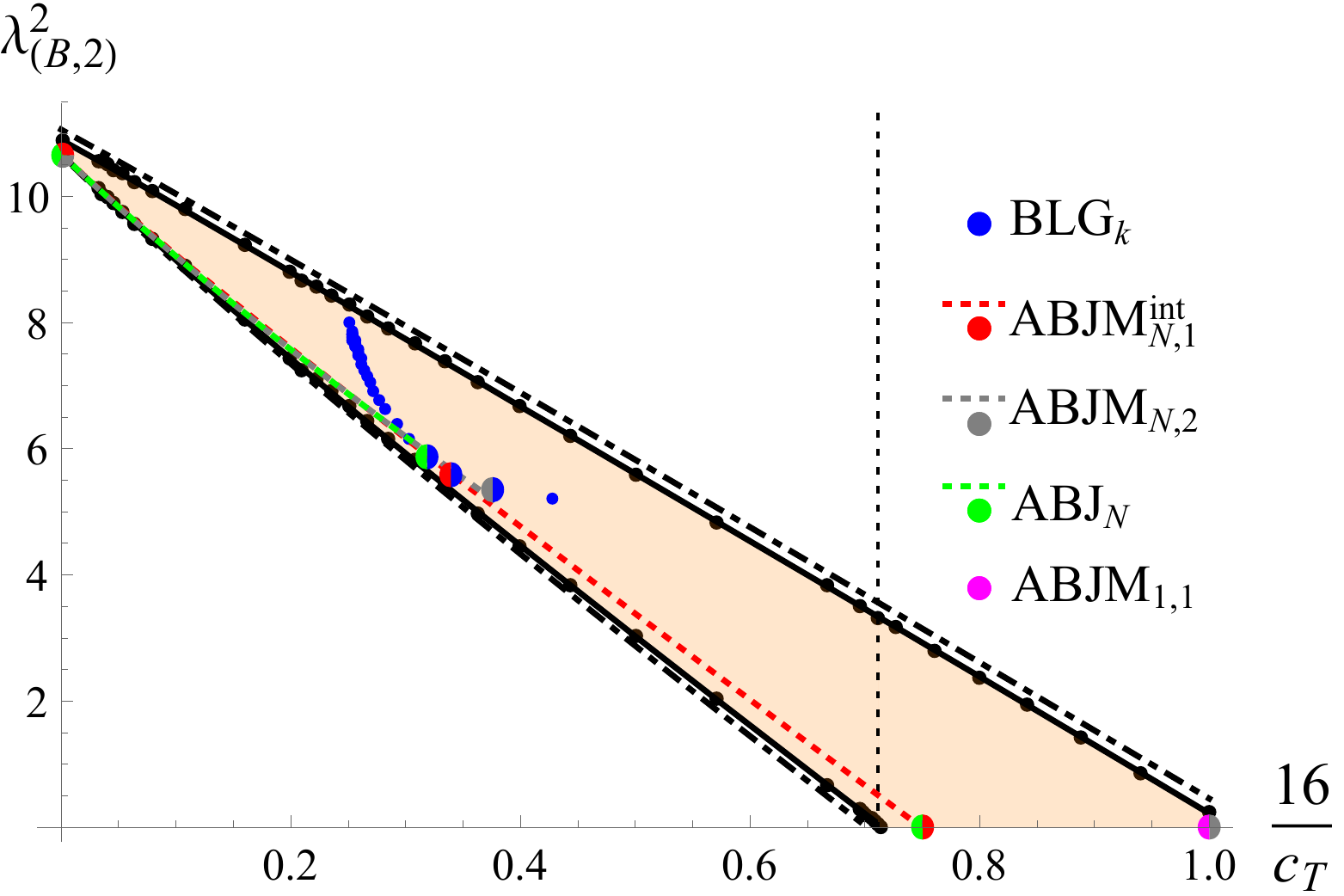}
 \caption{Upper and lower bounds on the $\lambda_{(B,2)}^2$ OPE coefficient in terms of the stress-tensor coefficient $c_T$, where the orange shaded region are allowed, and the plot ranges from the generalized free field theory limit $c_T\to\infty$ to the free theory $c_T=16$. The blue dots denote the exact values in Table \ref{valList} in BLG$_k$ for $k\geq1$. The magenta dot denotes the free ABJM$_{1,1}$ theory, the gray and green dots denote the exact values in Table \ref{valList} for ABJM$_{N,2}$ and ABJ$_{N}$, respectively, for $N=1,2,\infty$, and the red dots denote ABJM$_{N,1}^\text{int}$ for $N=2,3, \infty$. The red, gray, and green dotted lines show the large $N$ formulae \eqref{cTABJM} and \eqref{B2ABJM} for these theories for all $N\geq2$. The black dotted line denotes the numerical point $\frac{16}{c_T}\approx.71$ above which $\lambda_{(B,2)}^2=0$. The solid lines were computed with $\Lambda=43$. To show the level of convergence, the dashed lines are upper and lower bounds that were computed with $\Lambda=19$.}
\label{B22}
\end{center}
\end{figure}  

In Figure \ref{B22}, we show upper and lower bounds on $\lambda^2_{(B,2)}$ as a function of $\lambda^2_\text{Stress}/16=16/c_T$.  (The quantity $16/c_T$ ranges from $0$ (GFFT limit) to $1$ (free theory limit).)  We show our most accurate bounds with $\Lambda=43$ (solid line) as well as less accurate bounds with $\Lambda=19$ (dashed line), to show how converged the bounds are. The upper bounds seem to be converging at the same rate for all $c_T$, whereas the lower bounds seem more converged for larger $c_T$. The vertical dotted line shows the numerical point where $\lambda^2_{(B,2)}=0$. The red, gray, blue, and green dots denote some exact values listed in Table \ref{valList} for the interacting sector of ABJM$_{N,1}$, ABJM$_{N,2}$, BLG$_k$, and ABJ$_N$, respectively, where in all cases the dots go right to left for increasing $k,N$. We also list the free theory ABJM$_{1,1}$ as a magenta dot\footnote{ABJM$_{1,2}$ is not a free theory, but has the same stress tensor four-point function as a free theory.}. The red, gray, and green dotted lines show the large $N$ values for these theories for $N\geq2$ as given in \eqref{cTABJM} and \eqref{B2ABJM}.

There are several features of the BLG$_k$ plot that we would like to emphasize. For $k=1,2$, which are the values where BLG$_k$ theory is dual to a product theory (see footnote 6), the OPE coefficients lie in bulk of the allowed region. This is expected, because as described in \cite{Chester:2014fya}, all product theories generically lie in the bulk region. On the other hand, for $k=3,4$, which are the values where BLG$_k$ theory is dual to the interacting sector of ABJM$_{3,1}$ and ABJ$_2$, respectively, the OPE coefficients are close to saturating the lower bound. Lastly, for $k>4$, where it is not known whether the BLG$_k$ theories have an M-theory interpretation, the OPE coefficients of the BLG$_k$ theories interpolate between the lower and upper bounds. The $k\to\infty$ value is a little off from the upper bound, which is likely explained by the fact that the upper bound numerics are not fully converged. 

The ABJ(M) plot also has two interesting features. We first note that the ABJ$_1\cong\text{ABJM}_{2,1}^\text{int}$ theory is close to the numerical point where $\lambda^2_{(B,2)}=0$. In fact, this is the only known interacting theory with $\lambda^2_{(B,2)}=0$ \cite{Chester:2014mea}, so we suspect that with infinite accuracy the numerics would converge to this theory. We next note that all the ABJ(M) values seem to saturate the lower bound up to numerical error, with the exception of ABJM$_{2,2}$, which as explained before has the same stress tensor four-point function as a product theory and so must lie in the bulk.

\begin{figure}[t!]
\begin{center}
   \includegraphics[width=0.85\textwidth]{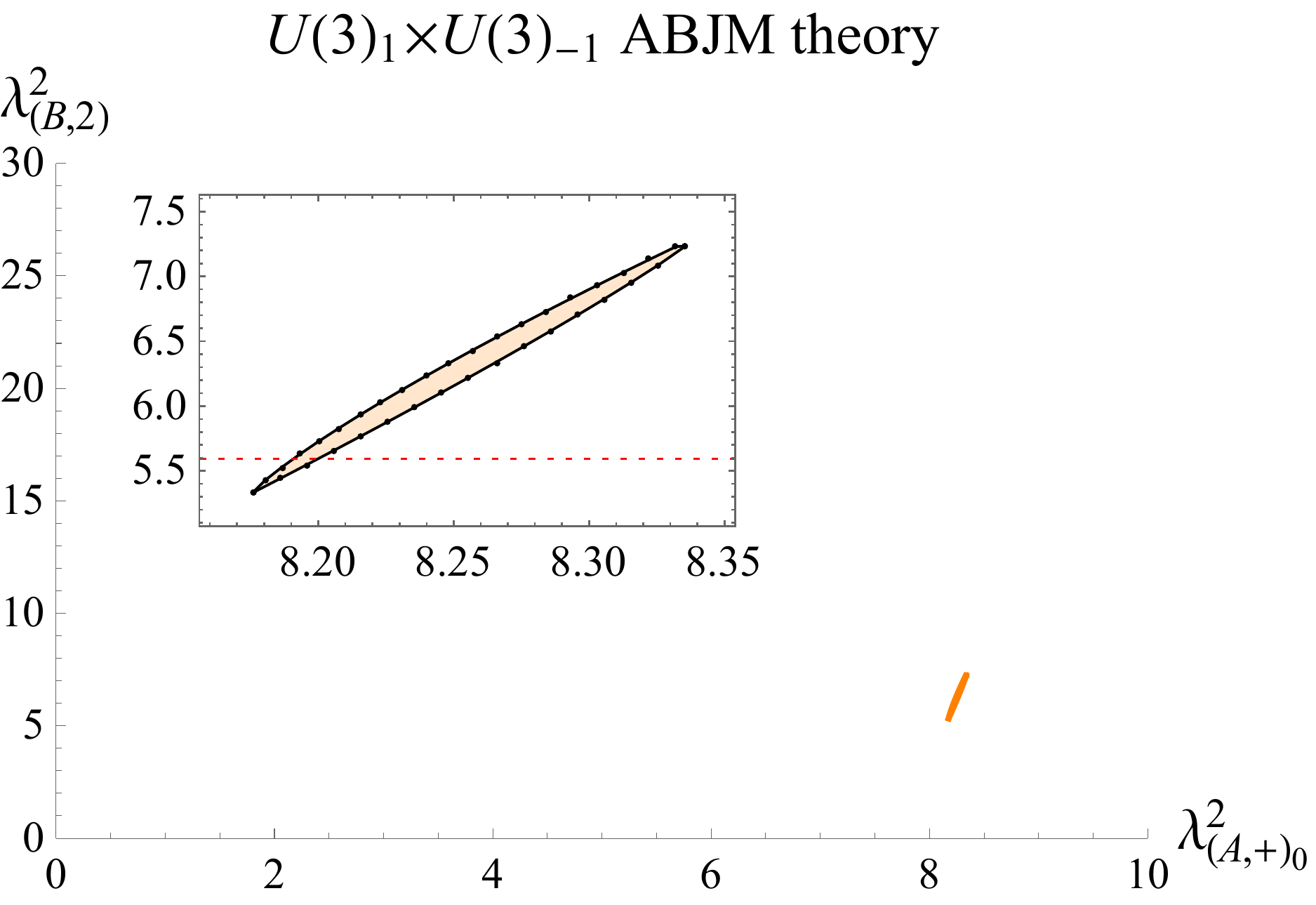}
 \caption{Bounds on $\lambda_{(B,2)}^2$ in terms of the $\lambda_{(A,+)_0}^2$ OPE coefficients at the ABJM$_{3,1}$ point with $\frac{16}{c_T}\approx0.340$. The orange shaded region is the allowed island, while the red dotted line shows the exactly known value given in Table \ref{valList} for $\lambda_{(B,2)}^2$ in this theory. These bounds were computed with $\Lambda=43$.}
\label{comp}
\end{center}
\end{figure}  

The fact that $\lambda^2_{(B,2)}$ for all unique ABJ(M) theories is close to saturating its lower bound may at first suggest that inputing any value of this OPE coefficient (within the bounds in Figure~\ref{B22}) into the numerical bootstrap code could uniquely specify that theory. To test this idea, in Figure~\ref{comp} we plot upper/lower bounds of $\lambda_{(B,2)}^2$ as a function of $\lambda_{(A,+)_0}^2$ at the ABJM$_{3,1}$ point with $\frac{16}{c_T}\approx0.340$ as given in Table \ref{valList}. While the allowed region is a small island, it does not shrink to a point. On the other hand, as the zoomed in plot shows, when $\lambda^2_{(B,2)}$ is at its extremal values then $\lambda_{(A,+)_0}^2$ is uniquely fixed. This matches the general numerical bootstrap expectation that all CFT data in the relevant four-point function is fixed at the boundary of an allowed region. Since the extremal value is very close to the exactly known value, as shown by the red dotted line, if we assume that it would exactly saturate the bound at infinite precision, then we can read off the spectrum of ABJM$_{3,1}$ by looking at the functional $\alpha$ that extremizes $\lambda_{(B,2)}^2$. Similar plots can be made for all the other unique ABJ(M) theories, so that $\lambda_{(B,2)}^2$ minimization gives the spectra of all theories with holographic duals that saturate the lower bound.

\section{Operator spectrum from numerical bootstrap}
\label{spec}

We now report our numerical results for the scaling dimensions and OPE coefficients of low-lying operators that appear in the OPE of $\cO_\text{Stress}$ with itself. We are interested in theories with holographic duals, and the lowest such known theories are ABJM$^\text{int}_{2,1}$ and $ \text{ABJ}_1$ with $\frac{16}{c_T}=.75$ and $\lambda^2_{(B,2)}=0$. As we see from Figure \ref{B22}, our numerics are not completely converged in that region, so we find that $\lambda^2_{(B,2)}=0$ at the numerical point $\frac{16}{c_T}\approx.71$. As such, in the following plots we will show results for $\frac{16}{c_T}>.71$.

Let us describe the $(A,0)$ unprotected operators that we expect to see in the spectrum. At the $c_T\to\infty$ generalized free field value we have the dimension $j+2+2n$ double trace operators $[\cO_\text{Stress}\cO_\text{Stress}]_{n,j}$ of the schematic form 
\es{dubtrace}{
[\cO_\text{Stress}\cO_\text{Stress}]_{n,j}=\cO_\text{Stress}\Box^n\partial_{\mu_1}\dots\partial_{\mu_j}\cO_\text{Stress}+\dots\,,
}
where $n=0,1,2,\dots$ and $\mu_i$ are space-time indices. The OPE coefficients of these operators  are given in Table \ref{Avalues}. At infinite $N$, these are the only operators with nonzero OPE coefficients. At large but finite $N$, there are also $m$-trace operators $[\cO_\text{Stress} ]^m_{n,j}$, with $m>1$, whose scaling dimension $\Delta^m_{n,j}$ and OPE coefficients $\lambda^m_{n,j}$ scale as \cite{Aharony:2016dwx}
\es{tracestuff}{
\Delta^m_{n,j}=&j+m +2n+O(1/c_T)\,,\qquad (\lambda^m_{n,j})^2=O(1/c_T^m)\,,
}
as well as single trace operators whose scaling dimension scales with $N$. For all ABJ(M) theories, $c_T\sim N^{3/2}$ \cite{Drukker:2010nc,Jafferis:2011zi} to leading order in large $N$, so the OPE coefficient squared of $m$-trace operators is suppressed as $N^{-3m/2}$. Even for the lowest trace operator after $[\cO_\text{Stress} ]^2_{n,j}$, i.e.~the triple trace operator $[\cO_\text{Stress} ]^3_{n,j}$, this suppression is extremely strong for even $N\sim10$. As a result, we do not expect the numerical bootstrap bounds to be sensitive to these higher trace operators at the currently feasible levels of precision. The situation is similar to high spin operators, which also have OPE coefficients that are highly suppressed \cite{Hogervorst:2013sma,Pappadopulo:2012jk, Rychkov:2015lca}, and so one can restrict to a finite number of operators with spin below some cutoff without affecting the numerics. It is the ability to ignore higher spin operator which in fact makes the numerical bootstrap possible at all.

For small $N$, we would expect the OPE coefficients of these higher trace operators to become large enough that they start to affect the numerics.  However, in this regime there is no clear distinction between higher trace and single trace operators because of trace relations.  Moreover, since the unprotected single trace operators are expected to have large scaling dimensions at large $N$, it is really not clear whether at small $N$ there should be an operator of small dimension that is continuously connected to the, say, triple trace operator at large $N$.

\subsection{$(A,0)$ scaling dimensions}

\begin{figure}[t!]
  \centering
\begin{center}
   \includegraphics[width=0.49\textwidth]{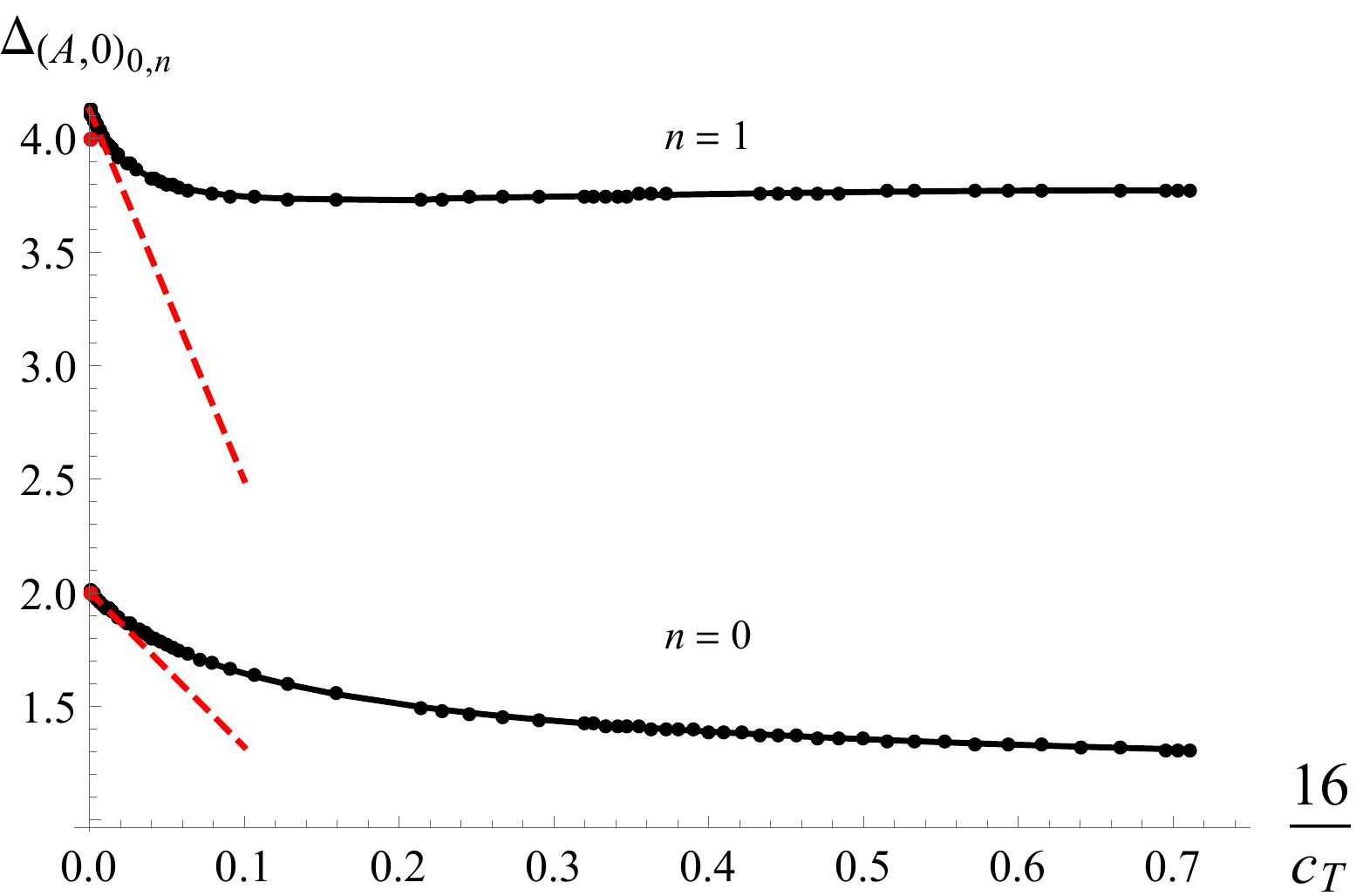}
    \includegraphics[width=0.5\textwidth]{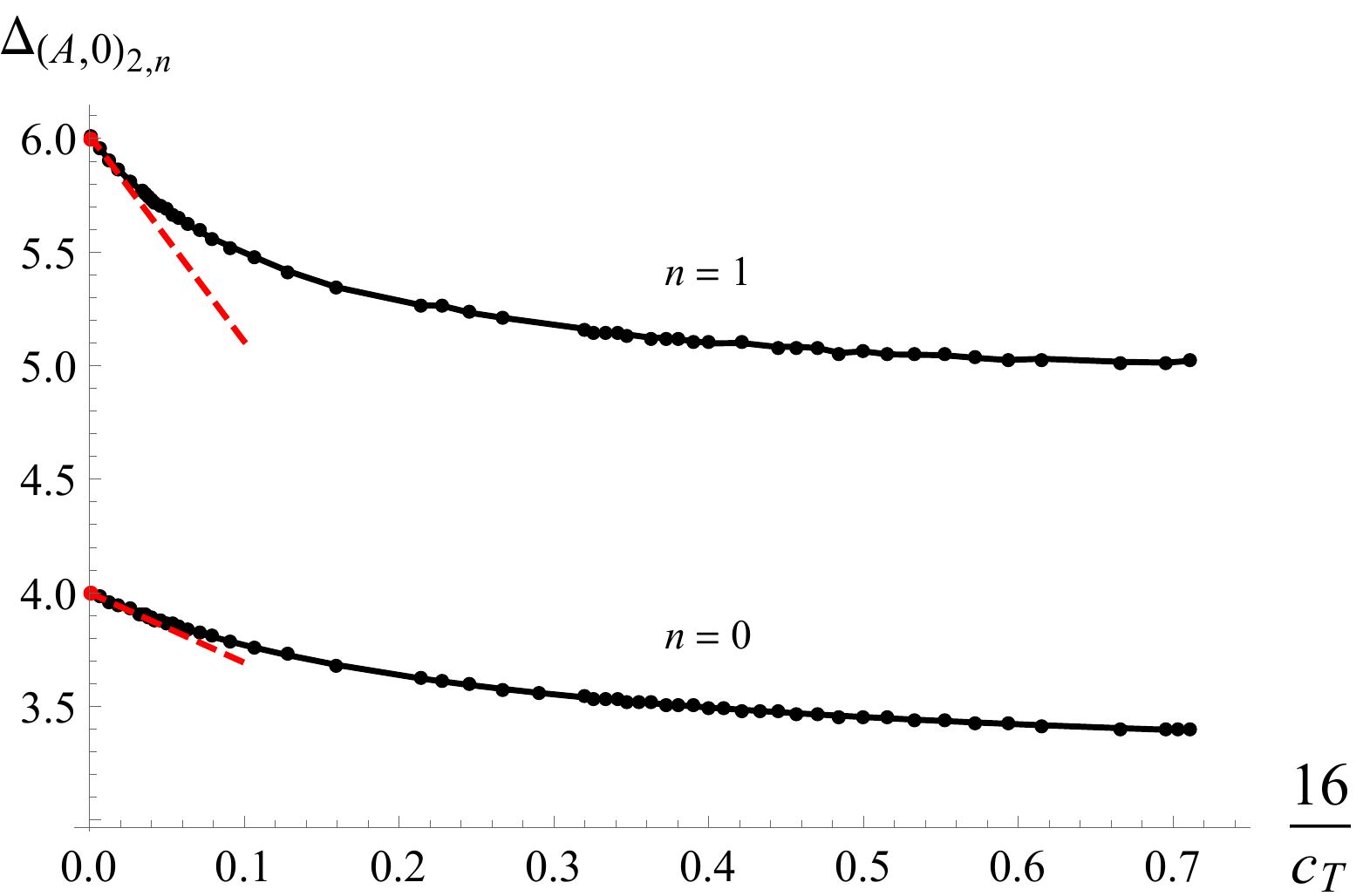}
    \includegraphics[width=0.5\textwidth]{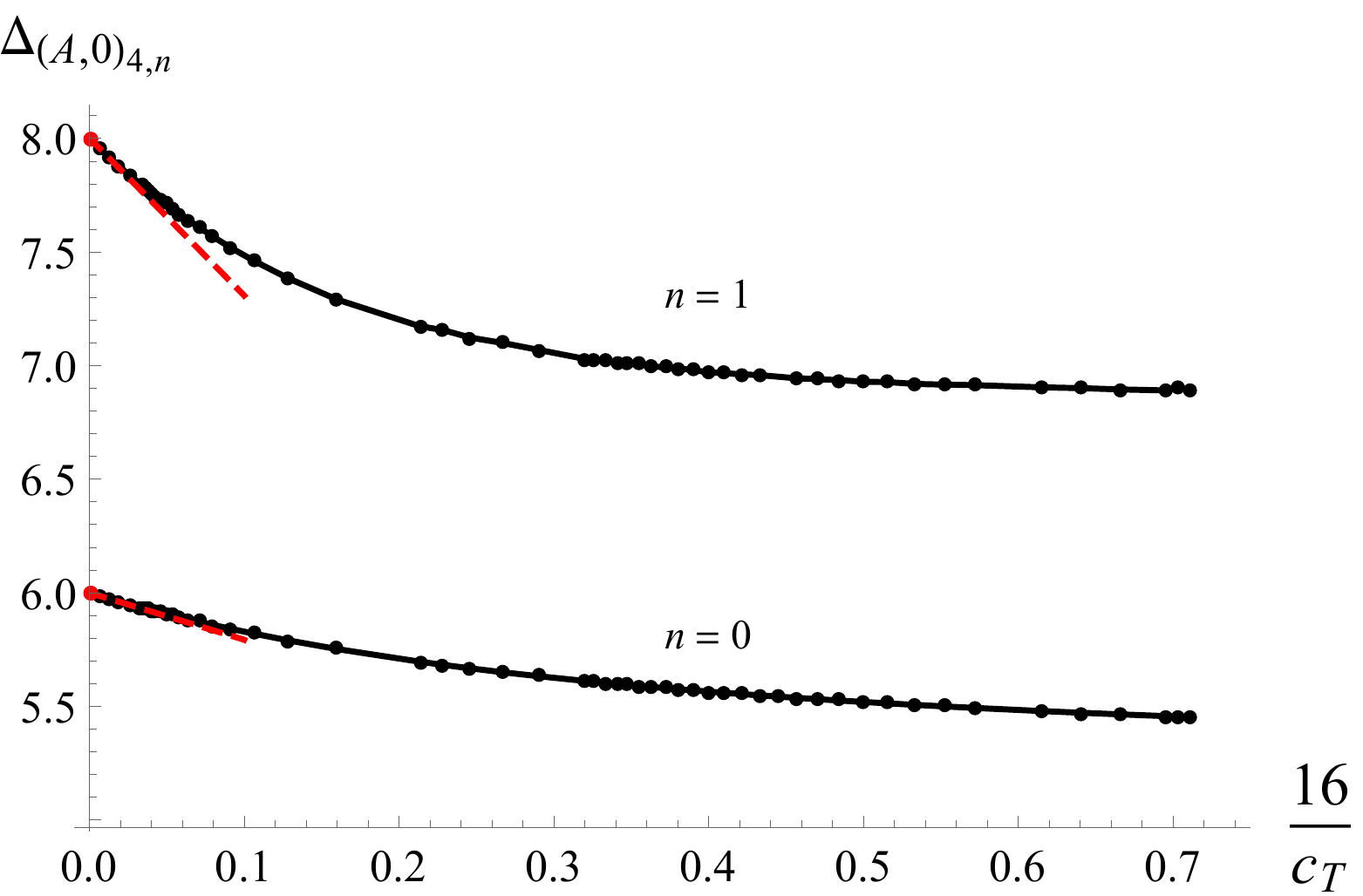}
\caption{
The scaling dimensions $\Delta_{(A,0)_{j,n}}$ for the two lowest $n=0,1$ long operators with spins $j=0,2,4$ in terms of the stress-tensor coefficient $c_T$, where the plot ranges from the generalized free field theory limit $c_T\to\infty$ to the numerical point $\frac{16}{c_T}\approx.71$ where $\lambda_{(B,2)}^2=0$. The red dots denote the known values $\Delta_j^{(n),\text{GFFT}}=j+2+2n$ for the generalized free field theory, while the red dotted lines show the linear fit for large $c_T$ given in \eqref{scalLarge}.  These bounds were computed with $\Lambda=43$. }
\label{longPlots}
\end{center}
\end{figure}

We can read off the scaling dimensions by looking at the zeros of the functional $\alpha(\Delta_{(A,0)_{j,n}})$ that minimizes $\lambda^2_{(B,2)}$. We trust those scaling dimensions that remain stable as we increase the number of derivatives $\Lambda$ in the bootstrap numerics. We observed that $\Delta_{(A,0)_{j,n}}$ for $j=0,2,4$ and $n=0,1$ are stable, and have values that in fact coincide with the upper bounds that we can independently compute for these quantities. 

In Figure \ref{longPlots} we show our numerical results for $\Delta_{(A,0)_{j,n}}$ for $n=0,1$ and $j=0,2,4$. All three of these plots show the same qualitative features. As described above, we only observe double trace operators, whose OPE coefficients are not suppressed at large $N$, i.e. small $c_T$. We can gauge how accurate these plots are by comparing to the $c_T\to\infty$ generalized free field values given in \eqref{superFree}. The plots seem to match the generalized free field theory values quite accurately. For large $c_T$, we find the following best fits
\es{scalLarge}{
\Delta_{(A,0)_{0,0}}&\approx 2.01-\frac{109}{c_T}\,,\qquad \Delta_{(A,0)_{2,0}}\approx4.13-\frac{49}{c_T}\,,\qquad \Delta_{(A,0)_{4,0}}\approx6.00-\frac{33}{c_T}\,,\\
 \Delta_{(A,0)_{0,1}}&\approx4.03-\frac{261}{c_T}\,,\qquad\Delta_{(A,0)_{2,1}}\approx6.02-\frac{145}{c_T}\,, \qquad \Delta_{(A,0)_{4,1}}\approx8.00-\frac{111}{c_T}\,.\\
}
As we see from Figure \ref{longPlots}, these linear fits are only accurate for large $c_T$.

\subsection{$(A,0)$, $(A,2)$, and $(A,+)$ OPE coefficients}

\begin{figure}[t!]
  \centering
\begin{center}
   \includegraphics[width=0.49\textwidth]{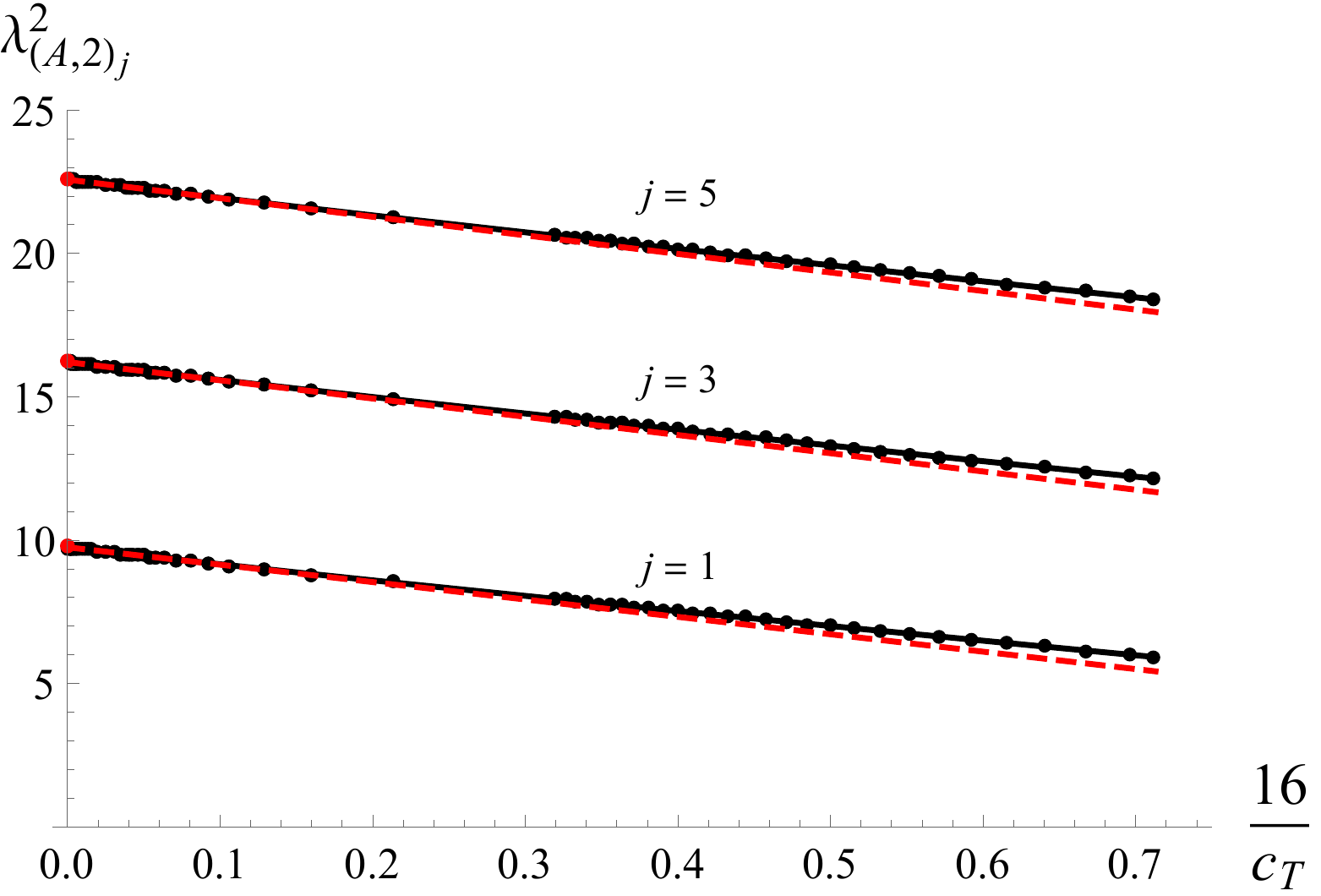}
    \includegraphics[width=0.5\textwidth]{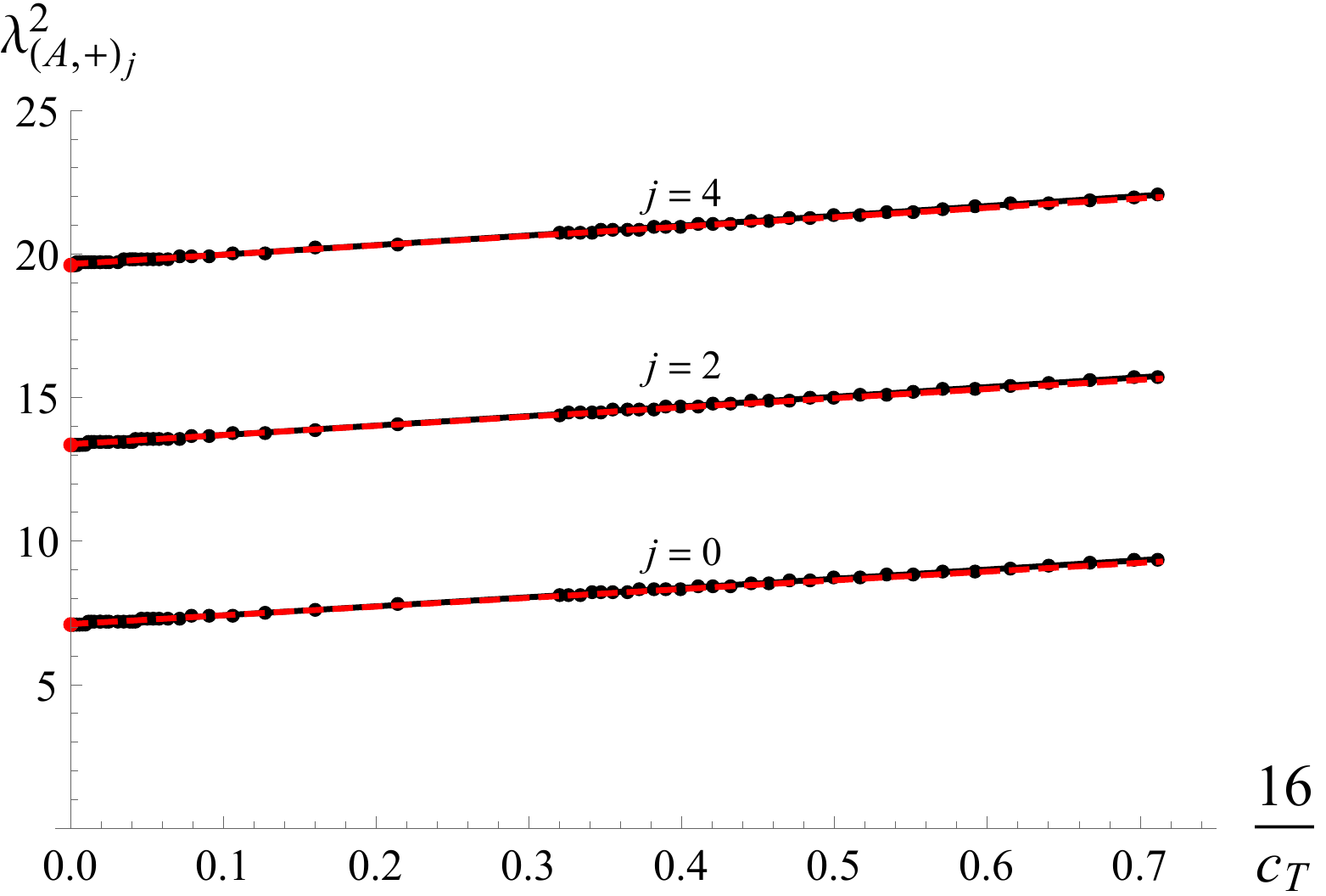}
    \caption{
The $\lambda_{(A,2)_j}^2$ and  $\lambda_{(A,+)_j}^2$ OPE coefficients with spins $j=1,3,5$ and $j=0,2,4$, respectively, in terms of the stress-tensor coefficient $c_T$, where the plot ranges from the generalized free field theory limit $c_T\to\infty$ to the numerical point $\frac{16}{c_T}\approx.71$ where $\lambda_{(B,2)}^2=0$. The red dots denotes denote the known values at the generalized free field theory points given in Table \ref{Avalues}, while the red dotted lines show the linear fit for large $c_T$ given in \eqref{semiLarge}. These bounds were computed with $\Lambda=43$.}
\label{APlots}
\end{center}
\end{figure}

Now that we have read off the low-lying scaling dimensions $\Delta_{(A,0)_{j,n}}$ from the extremal functional $\alpha$, we can compute low-lying OPE coefficients in the $(A,0)$, $(A,2)$, and $(A,+)$ multiplets by inputing $\Delta_{(A,0)_{j,n}}$ back into the bootstrap and computing upper and lower bounds on a given OPE coefficient. Since in the previous sections we only computed long multiplets with $n=0,1$, we will input the exact values for $n=0$ and then bound the continuum above the $n=1$ value, so that we can only extract long multiplet OPE coefficients with $n=0$. We find that the upper and lower bounds coincide, which matches our expectation that the extremal functional fixes these values. Note that in principle we could have extracted the OPE coefficients directly from $\alpha$ following the algorithm of \cite{ElShowk:2012hu,Simmons-Duffin:2016wlq}, but we found that this algorithm was very numerically unstable in our case.   

In Figure \ref{APlots} we show our numerical results for $\lambda_{(A,2)_j}^2$ and  $\lambda_{(A,+)_j}^2$ with $j=1,3,5$ and $j=0,2,4$, respectively. Just as with the $\Delta_{(A,0)_{j,n}}$ plots, these plots accurately match the generalized free field theory values listed in Table \ref{Avalues}. For large $c_T$, we find the following best fits
\es{semiLarge}{
\lambda^2_{(A,+)_0}&\approx7.11+\frac{49}{c_T}\,,\quad \lambda^2_{(A,+)_2}\approx13.37+\frac{51}{c_T}\,,\quad \lambda^2_{(A,+)_4}\approx19.65+\frac{52}{c_T}\,,\\
\lambda^2_{(A,2)_1}&\approx9.75-\frac{97}{c_T}\,,\quad \lambda^2_{(A,2)_3}\approx16.21-\frac{102}{c_T}\,,\quad \lambda^2_{(A,2)_5}\approx22.57-\frac{104}{c_T}\,.\\
}
As we see from Figure \ref{APlots}, these linear fits seem to be accurate for all values of $c_T$.

\begin{figure}[t!]
\begin{center}
   \includegraphics[width=0.85\textwidth]{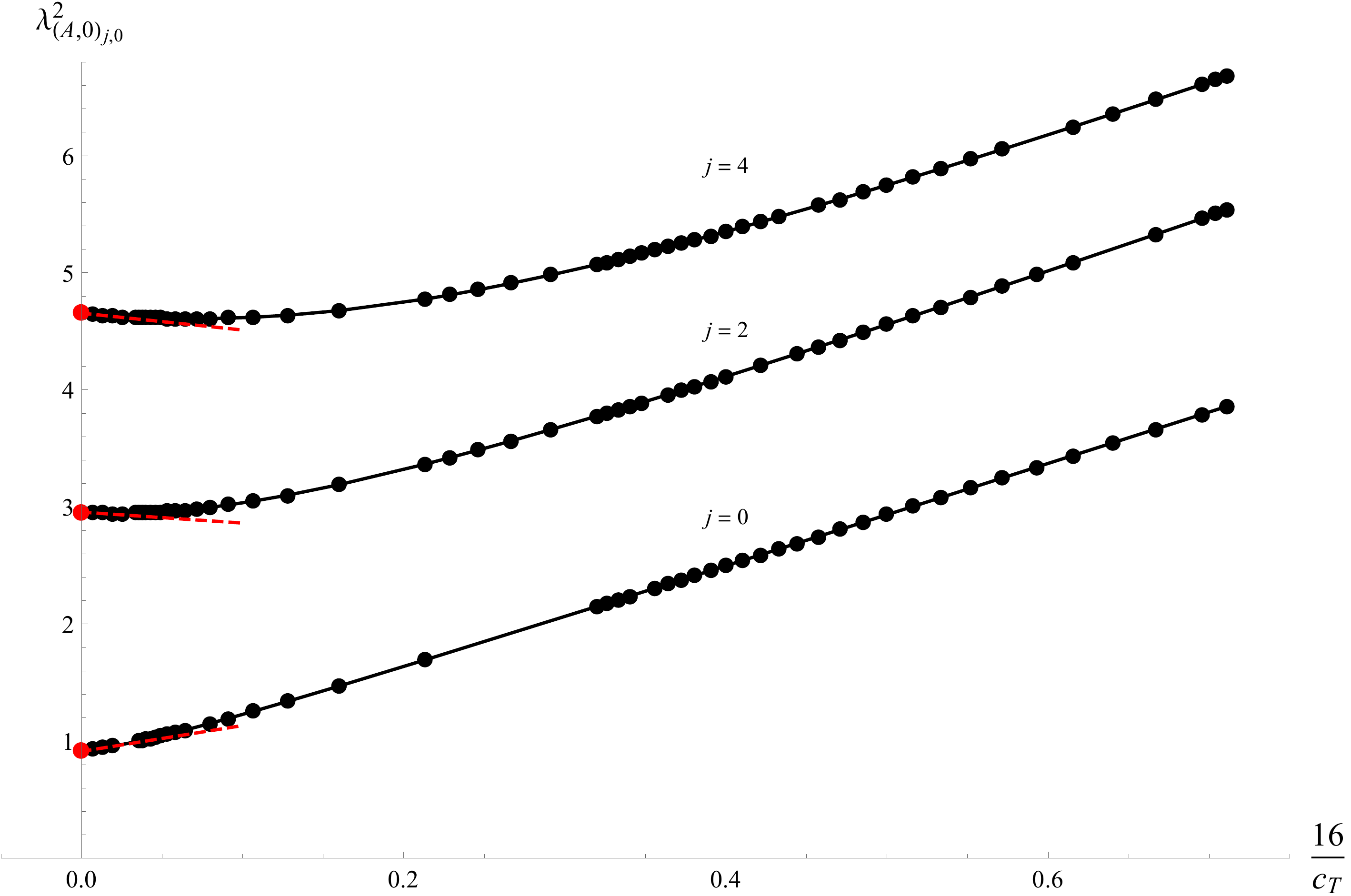}
 \caption{The $\lambda_{(A,0)}^2$ OPE coefficients for the three lowest spins in terms of the stress-tensor coefficient $c_T$, where the plot ranges from the generalized free field theory limit $c_T\to\infty$ to the numerical point $\frac{16}{c_T}\approx.71$ where $\lambda_{(B,2)}^2=0$. The red dots denotes denote the known values at the generalized free field theory points given in Table \ref{Avalues}, while the red dotted lines show the linear fit for large $c_T$ given in \eqref{longLarge}. These bounds were computed with $\Lambda=43$.}
\label{A0OPE}
\end{center}
\end{figure}   

In Figure \ref{A0OPE} we show our numerical results for $\lambda_{(A,0)_{j,n}}^2$ with $j=0,2,4$ and $n=0$. Just as with the $\Delta_{(A,0)_{j,n}}$ plots, these plots accurately match the generalized free field theory values listed in Table \ref{Avalues}. For large $c_T$, we find the following best fits
\es{longLarge}{
\lambda^2_{(A,0)_{0,0}}&\approx0.91+\frac{35}{c_T}\,,\qquad \lambda^2_{(A,0)_{2,0}}\approx2.96-\frac{15}{c_T}\,,\qquad \lambda^2_{(A,0)_{4,0}}\approx4.65-\frac{23}{c_T}\,.\\
}
As we see from Figure \ref{A0OPE}, these linear fits are only accurate for very large $c_T$.

\section{Discussion}
\label{disc}

There are two primary results in this work. Analytically, we have computed all half and quarter-BPS operator OPE coefficients that appear in the stress tensor four point function for all known $\mathcal{N}=8$ SCFTs.  Numerically, we have used the fact that these OPE coefficients for theories with M-theory duals saturate the numerical bootstrap bounds at large $c_T$ to extract all the low lying CFT data that appears in the stress tensor four point function. For smaller $c_T$, the numerics are not yet precise enough to determine which holographic theory saturates the bound, but we conjecture that at least one of the holographic theories does. These results are the first examples of scaling dimensions of unprotected operators and OPE coefficients in a large $N$ theory with an M-theory dual that have been computed for all values of $N$.

One notable feature of our numerical results is that we do not observe unprotected triple (or higher) trace operators in the spectrum, even though they should appear in $\cO_\text{Stress}\times \cO_\text{Stress}$ for finite $c_T$, i.e. finite $N$. We do not expect our numerics to be sensitive to these operators, because their OPE coefficients are highly suppressed, e.g.~the triple trace operator with dimension $\approx3$ at large $N$ has an OPE coefficient squared that is suppressed as $N^{-9/2}$. These operators are thus analogous to high spin operators, which also do not effect the numerical bootstrap because their OPE coefficients are highly suppressed \cite{Hogervorst:2013sma,Pappadopulo:2012jk, Rychkov:2015lca}.  However, with more precise numerics, we expect these operators to become visible, and it would be interesting to explore this point more in the future. It would also be interesting to observe whether the analogous triple trace operators also do not appear in the numerical bootstrap of other theories with large $N$ duals, such as the $\mathcal{N}=4$ superconfromal bootstrap in $d=4$ \cite{Beem:2013qxa,Beem:2016wfs}.

Another notable feature of our numerical results is that the values of the squared OPE coefficients of the double trace semi-short multiplets plotted in Figure~\ref{APlots} are approximately linear functions in $1/c_T$ for the entire range of $c_T$.  It would be interesting to find an explanation for this almost linearity.  From the bulk point of view, it implies that the only significant corrections to the $c_T = \infty$ values come from a tree level computation in the bulk.

There are several new directions that could strengthen our conjecture that only certain $\mathcal{N}=8$ theories with holographic duals saturate the numerical lower bounds.  From the numerical perspective, it would be useful to impose additional assumptions that would automatically exclude the theories that do not saturate the lower bounds.  For instance, in order to exclude the BLG$_k$ theories with $k>4$, one can apply the bootstrap to a mixed correlator between $\cO_\text{Stress}$ and the half BPS multiplet in $\mathfrak{so}(8)$ irrep $[0030]$. As one can check from the superconformal index, this latter operator does not exist for BLG$_k$ with $k>4$, while it does for generic ABJ(M) theories. Another feature of this mixed correlator is that the free multiplet appears in it, so by setting its OPE coefficient to zero one could also exclude the free theory. 

From the analytic perspective, it would be useful to include the non-perturbative corrections in $1/N$ to the results presented in this paper. These non-perturbative corrections have already been calculated for the $S^3$ free energy in many cases \cite{Marino:2009jd,Nosaka:2015iiw,Drukker:2010nc,Marino:2011eh,Hatsuda:2012dt,Hatsuda:2013gj,Calvo:2012du,Hatsuda:2013oxa,Kallen:2013qla,Honda:2013pea,Matsumoto:2013nya,Kallen:2014lsa,Codesido:2014oua}, but to extract the OPE coefficients from these results we would need an expression for these corrections as a smooth function of $m$. With these non-perturbative corrections included, we would have exact values of these OPE coefficients also for $\mathcal{N}=6$ ABJ(M) theories with gauge group $U(N)_k\times U(M)_{-k}$, so that we could see how these quantities interpolate between the $N,M\to\infty$ and fixed $k$ M-theory limit, the $N,M,k\to\infty$ and fixed $N/k$ Type IIA string theory limit, and the $N,k\to\infty$ and fixed $M$ higher spin theory \cite{Giombi:2011kc,Chang:2012kt} limit.

Lastly, it would also be interesting to calculate more BPS OPE coefficients in ABJ(M) theory in a large $N$ expansion using the Fermi gas approach \cite{Marino:2011eh}. For half and quarter-BPS operators that appear in $n$-point functions of the stress tensor, this could be done by taking more derivatives of the free energy as a function of the mass parameter $m$. For instance, three new OPE coefficients, $\lambda^{(B,2)}_{(B,2),(B,2)}$, $\lambda^{(B,+)}_{(B,+),(B,+)}$, and $\lambda^{(B,2)}_{(B,+),(B,+)}$,\footnote{$\lambda^{\text{Stress}}_{(B,+),(B,+)}$ also appears, but this OPE coefficient is related to $c_T$.} appear in the 6-point function, and crossing of the projection of this 6-point function to the 1d theory supplies two new constraints.\footnote{We thank Ran Yacoby for pointing this out to us.}  Thus, by taking 6 derivatives of the mass deformed $S^3$ partition function we can compute the integrated 6-point function in the 1d theory, and thereby determine all these OPE coefficients. For BPS operators that do not appear in any $n$-point functions of the stress tensor, such as operators in the $[00a0]$ irrep for odd $a$, we could still express their OPE coefficients as matrix integrals using the 1d methods of \cite{Dedushenko:2016jxl,Agmon:2017lga,Mezei:2017kmw}. These matrix integrals could then be computed as expectation values of $n$-body operators in the Fermi gas, along the lines of \cite{Klemm:2012ii}.

\subsection*{Acknowledgments}
\label{s:acks}

We thank Ofer Aharony, Simone Giombi, Igor Klebanov, Eric Perlmutter, David Simmons-Duffin, Yifang Wang, Ran Yacoby, Xi Yin, and Xinan Zhou for useful discussions.  We also thank the organizers and the participants of the bootstrap collaboration workshops at Stony Brook University and at the Simons Foundation, where part of this work was completed.  This work was supported in part by the Simons Foundation grant No.~488651.

\appendix

\section{Explicit formulas for $I_{n, k}$}
\label{blgDerivation}

These can be calculated exactly using the method from~\cite{Okuyama:2011su} by choosing an appropriate contour in the complex plane, applying Cauchy's theorem, and summing over residues. The integrals of interest to us are determined to be 
\es{Iints}{
I_{2, k}=&\begin{cases}
\frac{(-1)^{\frac{k-1}{2}}}{\pi}+\sum\limits_{s=1}^{k-1}(-1)^{s+1}\frac{k-2s}{2k^2}\tan\left[\frac{\pi s}{k}\right]^2\qquad\text{if $k$ is odd} \,,  \\
-\frac{(-1)^{\frac{k}{2}}}{\pi^2k}+\sum\limits_{\substack{s=1\\s\neq k/2}}^{k-1}(-1)^{s+1}\frac{(k-2s)^2}{4k^3}\tan\left[\frac{\pi s}{k}\right]^2\qquad\text{if $k$ is even} \,, \\
\end{cases}\\
I_{4, k}=&\begin{cases}
\frac{(-1)^{\frac{k+1}{2}}(3k^2-8)}{6\pi}+\sum\limits_{s=1}^{k-1}(-1)^{s}\frac{k-2s}{2k^2}\tan\left[\frac{\pi s}{k}\right]^4\qquad\text{if $k$ is odd} \,, \\
\frac{(-1)^{\frac{k}{2}}(k^2-8)}{6\pi^2k}+\sum\limits_{\substack{s=1\\s\neq k/2}}^{k-1}(-1)^{s}\frac{(k-2s)^2}{4k^3}\tan\left[\frac{\pi s}{k}\right]^4\qquad\text{if $k$ is even} \,, \\
\end{cases}\\
I_{6, k}=&\begin{cases}
\frac{(-1)^{\frac{k-1}{2}}(184-120k^2+25k^4)}{120\pi}+\sum\limits_{s=1}^{k-1}(-1)^{s+1}\frac{k-2s}{2k^2}\tan\left[\frac{\pi s}{k}\right]^6\qquad\text{if $k$ is odd} \,, \\
-\frac{(-1)^{\frac{k}{2}}(552-120k^2+7k^4)}{360\pi^2k}+\sum\limits_{\substack{s=1\\s\neq k/2}}^{k-1}(-1)^{s+1}\frac{(k-2s)^2}{4k^3}\tan\left[\frac{\pi s}{k}\right]^6\qquad\text{if $k$ is even}  \,.
\end{cases}
}
The quantities $I_{2, k}$ and $I_{4, k}$ had already been determined in \cite{Chester:2014fya}.

\section{Large $N$ formulae for $\lambda^2_{(B,2)}$}
\label{1dmore}

\es{B2ABJM}{
&\left(\lambda^{\text{ABJM}_{N,1}}_{(B,2)}\right)^2=\left(\frac{32}{3}\right)\left((112+45\pi^2)\text{Ai}[(N-3/8)(\pi^2/2)^{1/3}]+8(9+8\pi)(2\pi)^{2/3}\text{Ai}'[(N-3/8)(\pi^2/2)^{1/3}]\right)^{-2}\\
&\times\left[(-94976+8(3373+1080N+4800N^2+2560N^3)\pi^2+3465\pi^4)\text{Ai}\left[(N-3/8)({\pi^2}/{2})^{1/3}\right]^2\right.\\
&\left. \qquad+16(2\pi)^{2/3}(-5712-1664N+(981+872N)\pi^2) \text{Ai}\left[(N-3/8)({\pi^2}/{2})^{1/3}\right] \text{Ai}'\left[(N-3/8)({\pi^2}/{2})^{1/3}\right] \right.\\
&\left.\qquad-192(9+8N)^2(2\pi^2)^{1/3}\text{Ai}'\left[(N-3/8)({\pi^2}/{2})^{1/3}\right]^2
\right]\,,\\
&\left(\lambda^{\text{ABJM}_{N,2}}_{(B,2)}\right)^2=\left(\frac{32}{3}\right)\left(7\text{Ai}[(N-1/4)\pi^{2/3}]+4(1+2N)\pi^{2/3}\text{Ai}'[(N-1/4)\pi^{2/3}]\right)^{-2}\\
&\times\left[(-371+(58+120N^2+160N^3)\pi^2)\text{Ai}\left[(N-1/4){\pi}^{2/3}\right]^2\right.\\
&\left. \qquad+8\pi^{2/3}(-43-26N+(4+8N)\pi^2) \text{Ai}\left[(N-1/4){\pi}^{2/3}\right] \text{Ai}'\left[(N-1/4){\pi}^{2/3}\right] \right.\\
&\left.\qquad-24(1+2N)^2\pi^{4/3}\text{Ai}'\left[(N-1/4){\pi}^{2/3}\right]^2
\right]\,,\\
&\left(\lambda^{\text{ABJ}_{N}}_{(B,2)}\right)^2=\left(\frac{32}{3}\right)\left(7\text{Ai}[N\pi^{2/3}]+2(3+4N)\pi^{2/3}\text{Ai}'[N\pi^{2/3}]\right)^{-2}\\
&\times\left[(-371+(68+90N+240N^2+160N^3)\pi^2)\text{Ai}\left[N{\pi}^{2/3}\right]^2\right.\\
&\left. \qquad+4\pi^{2/3}(-99-52N+4(3+4N)\pi^2) \text{Ai}\left[N{\pi}^{2/3}\right] \text{Ai}'\left[N{\pi}^{2/3}\right] -6(3+4N)^2\pi^{4/3}\text{Ai}'\left[N{\pi}^{2/3}\right]^2\,
\right]\,.
}

\bibliographystyle{ssg}
\bibliography{Bootstrap}

\end{document}